\documentclass[12pt]{article}
\usepackage[T1]{fontenc}
\usepackage[dvips]{graphicx}
\graphicspath{{images/}}
\usepackage{verbatim}
\usepackage{amsmath,amsthm}
\usepackage{xspace}
\usepackage{pifont}
\usepackage{graphicx}
\usepackage{amssymb}
\usepackage{epic, eepic}
\usepackage{dsfont}
\usepackage{amssymb}
\usepackage{makeidx}
\usepackage{mathrsfs}
\usepackage{caption}
\usepackage{exscale}
\usepackage{color} 
\usepackage{overpic} 
\usepackage{bm}
\usepackage{bbm}
\usepackage{booktabs} 
\usepackage[hyphens]{url}
\usepackage{color, colortbl}
\usepackage{subcaption}
\usepackage[numbers]{natbib}
\RequirePackage[colorlinks,citecolor=blue,urlcolor=blue]{hyperref}
\definecolor{Gray}{gray}{0.9}
\usepackage{amsmath,afterpage}
\usepackage{epsf}
\usepackage{graphics,color}
\usepackage{authblk}
\usepackage[dvipsnames]{xcolor}
\usepackage{csquotes}
\usepackage{pgfplots}
\usepackage{rotating}
\usepackage{pdflscape}
\usepackage{booktabs}
\usepackage{tikz}
\usepackage{array}
\usepackage{arydshln}
\usepackage{breakcites}
\usepackage{breakurl}
\usepackage{afterpage}

\usetikzlibrary{backgrounds,calc}

\pgfkeys{/tikz/.cd,
  timespan/.store in=\timespan,
  timespan= Week,
  timeline width/.store in=\timelinewidth,
  timeline width=30,
  timeline height/.store in=\timelineheight,
  timeline height=1,
  timeline offset/.store in=\timelineoffset,
  timeline offset=0.15,
  initial week/.store in=\initialweek,
  initial week=1,
  end week/.store in=\endweek,
  end week=2,
  time point/.store in=\timepoint,
  time point=0.5,
  between day/.style args={#1 and #2 in #3}{
    initial week=#1,
    end week=#2,
    time point=#3,
  },
  between week/.style args={#1 and #2 in #3}{
    initial week=#1,
    end week=#2,
    time point=#3,
  },
  between month/.style args={#1 and #2 in #3}{
    initial week=#1,
    end week=#2,
    time point=#3,
  },
  between year/.style args={#1 and #2 in #3}{
    initial week=#1,
    end week=#2,
    time point=#3,
  },
  involvement degree/.store in=\involvdegree,
  involvement degree=2cm,
  phase color/.store in=\phasecol,
  phase color=red!50!orange,
  phase appearance/.style={
    circle,
    opacity=0.9,
    minimum size=\involvdegree,
    fill=\phasecol
  },
}
\newif\ifcustominterval
\pgfkeys{/tikz/timeline/.cd,
  custom interval/.is if=custominterval,
  custom interval=false,
}

\pgfkeys{/tikz/milestone/.cd,
  at/.store in=\msstartpoint,
  at=phase-1.north,
  circle radius/.store in=\milestonecircleradius,
  circle radius=0.1cm,
  direction/.store in=\msdirection,
  direction=90:2cm,
  text/.store in=\mstext,
  text={},
  text options/.code={\tikzset{#1}},
}

\newcommand{\timeline}[2][]{
  \pgfkeys{/tikz/timeline/.cd,#1}
  \draw[fill=white,opacity=0.95] (0,0) rectangle (\timelinewidth,\timelineheight);

  \ifcustominterval%
    \foreach \smitem [count=\tlmxi] in {#2}  {\global\let\maxsmitem\tlmxi}%
  \else%
    \foreach \smitem [count=\tlmxi] in {1,...,#2}  {\global\let\maxsmitem\tlmxi}%
  \fi%
  
  \pgfmathsetmacro\position{\timelinewidth/(\maxsmitem+1)}
  \node at (0,0.5\timelineheight)(\timespan-0){\phantom{Week 0}};
 
  \ifcustominterval%
    \foreach \x[count=\tlmxi] in {#2}{%
      \node[text=black,text depth=0pt]at +(\tlmxi*\position,0.5\timelineheight) (\timespan-\tlmxi) {\timespan\ \x};%
    }%
  \else%
    \foreach \x[count=\tlmxi] in {1,...,#2}{%
      \node[text=white, text depth=0pt]at +(\tlmxi*\position,0.5\timelineheight) (\timespan-\tlmxi) {\timespan\ \x};%
    }%
  \fi%
}

\newcounter{involv}
\newcommand{\phase}[1]{
\stepcounter{involv}
\node[phase appearance,#1] 
 (phase-\theinvolv)
 at ($(\timespan-\initialweek)!\timepoint!(\timespan-\endweek)$){};
}

\newenvironment{phases}{\begin{pgfonlayer}{background}}{\end{pgfonlayer}}

\newcommand{\addmilestone}[1]{
\pgfkeys{/tikz/milestone/.cd,#1}
\draw[double,fill] (\msstartpoint) circle [radius=\milestonecircleradius];
\draw(\msstartpoint)--++(\msdirection)node[/tikz/milestone/text options]{\mstext};
}

\usepackage[section]{algorithm}

\numberwithin{equation}{section}

\newcommand{\bd}{\begin{displaymath}}
\newcommand{\ed}{\end{displaymath}}
\newcommand{\be}{\begin{equation}}
\newcommand{\ee}{\end{equation}}
\newcommand{\bq}{\begin{eqnarray}}
\newcommand{\eq}{\end{eqnarray}}
\newcommand{\bn}{\begin{eqnarray*}}
\newcommand{\en}{\end{eqnarray*}}

\title{Evidence of Crowding on Russell 3000 Reconstitution Events}
\author{Alessandro Micheli\thanks{AM is supported by the EPSRC Centre for Doctoral Training in Mathematics of Random Systems: Analysis, Modelling and Simulation (EP/S023925/1)}}
\author{Eyal Neuman }
\affil{Department of Mathematics, Imperial College London}
 
\begin{document}

 \vspace{-0.5cm}
\maketitle

\begin{abstract}
We develop a methodology which replicates in great accuracy the FTSE Russell indexes reconstitutions, including the quarterly rebalancings due to new initial public offerings (IPOs). 
While using only data available in the CRSP US Stock database for our index reconstruction, we demonstrate the accuracy of this methodology by comparing it to the original Russell US indexes for the time period between 1989 to 2019. A python package that generates the replicated indexes is also provided \cite{micheli_2020}. 

As an application, we use our index reconstruction protocol to compute the permanent and temporary price impact on the Russell 3000 annual additions and deletions, and on the quarterly additions of new IPOs . We find that the index portfolios following the Russell 3000 index and rebalanced on an annual basis are overall more crowded than those following the index on a quarterly basis. This phenomenon implies that transaction costs of indexing strategies could be significantly reduced by buying new IPOs additions in proximity to quarterly rebalance dates.

\end{abstract} 


\begin{description}
\item[Keywords:] crowding, indexing strategies, price impact, Russell Index.
\end{description}

\bigskip

\section{Introduction}
FTSE Russell is, quoting the company web-page \cite{ftse_home}, a ``\emph{global provider of benchmarks, analytics, and data solutions with multi-asset capabilities}''.  The company maintains a wide range of indexes varying for geographic regions, weighting procedures and asset classes.

In US markets, FTSE Russell most prominent products are the \textit{Russell US indexes}: the Russell 1000, 2000, 3000 and 3000E indexes track rosters of US companies across different market capitalizations. Part of the strength of Russell US indexes resides in their modularity. As shown in Table \ref{table:russell_modules}, each index is composed according to different investment styles, therefore offering an extended and meticulous coverage for the US Equity market. For example, the Russell 3000 measures the performance of the 3,000 largest public companies in the US by total market capitalization and represents approximately 98 percent of the American public equity market. On the other hand, the Russell 1000 Defensive Index is much more specialized as it includes those Russell 1000 Index companies that are more stable and are less sensitive to economic cycles, credit cycles and market volatility.

Such indexes are often used by portfolio managers as benchmarks for US equity market performances across different market segments. It does not come as a surprise then, that Russell US indexes are the go-to equity universe for a wide body of academic literature, including portfolio management research \cite{Cremers2020,Biktimirov2004,Boone2015,Chang2014,Chen2005} as well as market microstructure e.g. \cite{bucci2019slow,volpati2020zooming,Capponi2019,Zarinelli2015,bucci2019trading,Bormetti2015,Bucci2020,Lillo-Calcagnile}.  

The rosters of securities in the Russell U.S. indexes have also received attention for the presence of the so called \textit{``index effects''.} It has been empirically observed that the securities added to equity indexes receive positive returns concurrently with their index additions and shortly thereafter. The main indexes on which such effects are observed are the S\&P500 and the Russell U.S. indexes, with many studies, such as \cite{Madhavan2001,Cai2008,Chen2006,Chang2014,Petajisto2011}, providing evidences in support of the existence of the aforementioned abnormal returns. 

As for the Russell U.S. indexes, Madhavan \cite{Madhavan2001} first analyzed the presence of statistically significant abnormal returns attributable to the annual reconstitution of Russell 2000 and Russell 3000 indexes. Moreover, Madhavan explained the abnormal returns due to microstructure effects such as price pressure and changes in liquidity. The mechanisms generating these abnormal returns phenomena were further investigated and tested by Chen in \cite{Chen2006}.  Cai and Todd Houge in \cite{Cai2008} compared the performance of a buy-and-hold strategy of the Russell 2000 index to the returns of a portfolio following the annually rebalanced Russell 2000 index. The latter was shown to be significantly more profitable in a time scale of 5 years.  More recently, Onayev and Zdorovtsov \cite{Onayev2008} have found evidence of strategic predatory trading behaviour around the annual reconstitution, whereby closing prices of companies are manipulated in order to influence their index membership.

\begingroup
\renewcommand{\arraystretch}{1.25} 
\begin{table*}[h!]
\centering
\begin{small}
\begin{tabular}{@{}p{4.8cm}p{4.8cm}p{4.8cm}@{}  }
\toprule
\multicolumn{3}{c}{ \textbf{Russell U.S. Indexes}} \\
\toprule
\textbf{Broad market} & \textbf{Large cap}  & \textbf{Small cap}  \\ \midrule
Russell 3000E Index  & Russell 1000 Index & Russell 2000 Index \\ 
Russell 3000E Value Index  & Russell 1000 Value Index & Russell 2000 Value Index \\ 
Russell 3000E Growth Index  & Russell 1000 Growth Index & Russell 2000 Growth Index \\ 
Russell 3000 Index  & Russell 1000 Defensive Index & Russell 2000 Defensive Index \\ 
Russell 3000 Value Index  & Russell 1000 Dynamic Index & Russell 2000 Dynamic Index \\ 
Russell 3000 Growth Index  & Russell 1000 Growth-Defensive Index & Russell 2000 Growth-Defensive Index \\ 
Russell 3000 Defensive Index  & Russell 1000 Growth-Dynamic Index & Russell 2000 Growth-Dynamic Index \\ 
Russell 3000 Dynamic Index  & Russell 1000 Value-Defensive Index & Russell 2000 Value-Defensive Index \\ 
Russell 3000 Growth-Defensive Index  & Russell 1000 Value-Dynamic Index & Russell 2000 Value-Dynamic Index \\ 
Russell 3000 Growth-Dynamic Index &  &  \\ 
Russell 3000 Value-Defensive Index & &  \\ 
Russell 3000 Value-Dynamic Index  & &  \\ 
\bottomrule
\end{tabular}
\end{small}
\vspace{\abovecaptionskip}
\caption{Russell US indexes by investment style and market sector. Table originally published in Section ``Construction and Methodology'' of  \cite{ftse_russell_2020}.}
\label{table:russell_modules}
\end{table*}
\endgroup

One of the main features distinguishing the Russell U.S. indexes across all others US stocks equity indexes is their rebalance procedure. In general, rebalance procedure of equity indexes are not necessarily publicly disclosed and sometimes presents  some degree of arbitrariness.
 For example, as discussed in \cite{Cai2008,Petajisto2011}, Standard \& Poor's maintains a proprietary selection process used to discern which stocks will belong to the new issue of the index and make adjustments whenever it considers it to be necessary. Nonetheless, even though such procedures remain undisclosed, the S\&P500 historical constituent securities are available to researchers via the WRDS database maintained by the Wharton School of the University of Pennsylvania.  
 
On the other hand, we have the FTSE company that implements a publicly available fully deterministic rebalance algorithm for its Russell indexes, but which prefers not to publicly disclose their historical index compositions. Bloomberg L.P. terminals offer the list of the companies in the indexes but neither the constituents securities nor the index weights are available.

The FTSE index compositions are available to buy for financial institutions and funds. On the academic side, the WRDS database 
has recently started providing a Russell index historical dataset for 21 indexes, with a substantial annual fee. However this dataset provides only information on index weights and companies contributions to returns. It falls short of more refined information such as quarterly and annual ranking and rebalance days, and on historical lists of securities for companies which are traded across different classes of shares. See Section \ref{sec-meth} for a detailed description of these features and their importance to the index reconstitution. Since these features play a crucial role on indexes reconstitution, tracking them in a consistent framework is important for academic research (see e.g. the analysis in Section \ref{sec-impact}). There seems to be a gap in Russell data for research purposes, in the sense that there is no recognised source from which academic researchers and financial institutions' internal researchers can borrow detailed Russell US indexes data from. Therefore, it is very likely that the notion and composition which is used to approximate the Russell US indexes, e.g. the Russell 3000 or Russell 2000 indexes, could be diversified across different academic papers.

Agreeing upon a common \textit{``definition''} of what are Russell indexes could benefit the financial research community as a whole and is one of the main goals of this work. On some very elementary scientific grounding, sharing the notion of initial data, common to all financial data analyses, allows for a higher degree of reproducibility of the results. In the first part of this paper (Sections \ref{sec-meth}--\ref{sec-res}) we develop a methodology which is based only on data available in the CRSP database of the Wharton Research Data Services (see Section \ref{sec-data} for more details on the database). This methodology allows us to replicate in great accuracy the Russell 1000, 2000 and 3000 indexes weights and returns, and to track new additions and deletions. We demonstrate the accuracy of this methodology by comparing our suggested reconstitution procedure versus the original Russell US indexes for the time period June 1989 to June 2019 (see Section \ref{sec-res}). A python package named \emph{pyndex} that generates the indexes according to our methodology is also provided in \cite{micheli_2020}.

The impact of sharing such initial data might vary across different studies, ranging from marginal to impactful, nonetheless we still remark that it could only be beneficial and would bring the research community a step closer to the conclusion of disputes on results based on the quality of the data being analysed. Similarly to the natural sciences, we remark that the validity of quantitative claims is settled only by referring to the observations of the phenomenon, which, by nature, strongly depend on the data analysed. 


As a first application for our index reconstruction methodology, we study crowding on indexing strategies around Russell 3000 reconstitution events, which starting from 2004 occurs every quarter. Before we describe our analysis  on this topic, we survey some existing literature on crowding in financial markets.

Over the last 20 years, the phenomenon of \textit{crowding} in financial markets has increasingly gained attention both from academics as well as from financial institutions. It is a subject of many research works studying both theoretical and empirical aspects including \cite{Cont2000,volpati2020zooming,Bucci2020,Barroso2017,Caccioli2015,Caccioli2014,Khandani2008}.  

Crowding is often considered to be an explanation for sub-par performances of investments as well the development of systemic risk in financial markets. The presence of largely overlapping portfolios comes at the expense of portfolio managers, also in terms of transaction costs, as affine positions usually lead to similar trades. 

Cont and Bouchaud \cite{Cont2000} proposed a simple mathematical model in which the communication structure between agents gives rise to heavy tailed distribution for stock returns. This established a theoretical connection between crowding and stock markets shortfall. The aforementioned portfolios overlap was shown to be a considerable factor in the August 2007 Quant Meltdown. Using simulated returns of overlapping equity portfolios, Khandani and Lo \cite{Khandani2008} showed that combined effects of portfolio deleveraging following by a temporary withdrawal of market-making risk capital was one of the main drivers of the 2007 Quant Meltdown. Caccioli et al. \cite{Caccioli2015,Caccioli2014} developed a mathematical model for a network of different banks holding overlapping portfolios. They investigated the circumstances under which systemic instabilities may occur as a result of various parameters, such as market crowding and market impact.  Recently, Volpati et al. \cite{volpati2020zooming} measured significant levels of crowding in U.S. equity markets for Momentum signals  as well as for Fama-French factors signals, even though with smaller significance.

As already mentioned, we apply our index reconstruction methodology in order to measure crowding effect on Russell indexes around reconstitutions events. It was reported by Madhavan \cite{Madhavan2001} and others, that around annual index reconstitution events, there are significant abnormal returns on stocks which are new additions or deletions from the index. 
These returns are caused by portfolio trading strategies which anticipate the change in the stock price for new additions or deletions.
The returns of these stocks are typically decomposed into two parts. The first is called temporary price impact, and it  describes the returns that revert back within one month from index rebalance day. The permanent price impact captures sustainable stock returns, which accumulates within two months from the index reconstitution announcement (see more details in Section \ref{sec-impact}). It follows that the reconstitution events of the Russell 3000 index serve as prominent examples for studying crowding effects of trading strategies.

Starting from 2004, the Russell indexes received quarterly \textit{additions} to take into account the changes brought in the market by newly listed securities, that is IPOs, which took place between annual rebalancings (see additional details in Section \ref{sec-meth}). The practice of quarterly updates to the indexes was continued ever since and is currently still in use. To our knowledge, none of the papers that studied the Russell index reconstitution effect has dealt with these quarterly additions. The reconstruction methodology, which is developed in this paper, can assist us to preform a more refined analysis of crowding phenomenon on Russell 3000 index around reconstitution dates. This is the second objective of this paper.

In Section \ref{sec-impact} we compute the permanent and temporary price impact on the Russell 3000 stock additions and deletions, using the annual index portfolios generated with our protocol. We also track the quarterly additions to the  Russell 3000 index. We find that the price impacts of the aforementioned quarterly additions are overall compatible with the hypothesis that the majority of market participants track the Russell 3000 index on an annual basis rather than on a quarterly basis. Such findings are consistent with the belief that the portfolio strategies following the Russell 3000 index rebalance on an annual basis are more crowded than those following the Russell 3000 index rebalance on a quarterly basis.

This paper is structured as follows. In Section \ref{sec-meth} we describe the precise methodology of the  FTSE Russell indexes reconstitution which includes the quarterly rebalancings due to new initial public offerings (IPOs). In Section \ref{sec-data} we describe the data that we are using in order to reconstruct the indexes in this paper. Section \ref{sec-res} is dedicated to our methodology which approximates the Russell US indexes to our results which replicate the indexes. In Section \ref{sec-impact} we determine the temporary and permanent price impact generated by the annual index additions as well as examine the existence of crowded trades around the annual Russell 3000 reconstitution.  In Section \ref{sec-con} we present the conclusions of this paper.

\section{The FTSE Russell indexes reconstitution Methodology} \label{sec-meth} 
In this section we describe the main features in the original FTSE Russell indexes reconstitution methodology.

The FTSE Russell US 1000, 2000 and 3000 are equity capitalisation-weighted indexes that currently follow an annual rebalance procedure, which was first adopted in June 1989. 
As further discussed in \cite{Cai2008}, the indexes followed a quarterly rebalance schedule from  1979 to 1986 and a semi-annual one from 1987 to 1989.
In the 2004 rebalance calendar, the Russell indexes received quarterly \textit{additions} to take into account the changes brought in the market by newly listed securities, that is IPOs which took place between annual rebalancings. The practice of quarterly updates to the indexes was continued ever since and is currently still in use. 

 We remark that the newly issued securities added at each quarter do not replace any other company already in the indexes, in fact quoting the 2004 press release \cite{business_wire_2004} of Russell Investments\footnote{Russell Investments controlled the Russell US indexes until its index division was bought by LSE Group in 2015 and subsequently renamed FTSE Russell.}:

\begin{displayquote}
 \textit{``As IPOs are added to Russell indexes each quarter, Russell will not delete existing index members to make room for them, but will continue to reconstitute the indexes fully each year at the end of the second quarter.''}
\end{displayquote}

Inclusion in the Russell indexes is established systematically via a set of rules which we will briefly summarise, without intending to be fully exhaustive. For the full list of the current selection rules we invite the reader to consider the official documentation available at \cite{ftse_russell_2020}.  

Each year on the \textit{rank day}, which takes place in May,  all U.S.-domiciled companies with stock prices greater than \$1.00 are ranked according to their market capitalisation. The total market capitalisation of a company is computed by determining the shares of common stock, non-restricted exchangeable shares and partnership units/membership interests while excluding any other form of shares, such as convertible preferred stocks, foreign securities as well as American Depositary Receipts (ADR). Explicitly, as discussed in \cite{ftse_russell_2020}, exchangeable shares are shares which may be exchanged, on a one-for-one basis, at the owner's option at any time, while membership interests or partnership units embody an economic interest in a limited liability company or limited partnership. 
\par For a company which is traded across different classes of shares, e.g. Berkshire Hathaway, FTSE Russell first determines its so called \textit{pricing vehicle}: the share class with the highest two-year trading volume as of the corresponding rank day in May. Hence, the total market capitalisation is computed by multiplying the  cumulative sum of shares across all classes by the close price of the pricing vehicle on the rank day. Only companies with a total market capitalisation higher than 30 million U.S. dollars are included in the ranking of the Russell US indexes.

Once the ranking has been established, the 3000 companies with the highest market capitalisation fall in the Russell 3000 index. The top 1000 companies in the Russell 3000 index in turn, constitute the Russell 1000 index, while the bottom 2000 determine the Russell 2000 index. The top 4000 companies in the ranking with total market capitalisation higher than 30 million U.S. dollars, or all the available securities in case they are less than 4000, constitute the Russell 3000E.   
 
The weight corresponding to each security admitted to the index is computed as follows. The outstanding shares of a security are adjusted to only include the number of available shares which can be traded by the public, the so called \textit{``free float''}. In fact, it is possible that some of the shares, as those held by government or other third party, might not be available for trading. The adjustment is done based on information contained in governmental filings, such as those submitted to the Securities and Exchange Commission (SEC). Further information about the float-adjustment procedure for the rebalance year 2020 can be found in Section ``Methodology Enhancements'' of \cite{ftse_calendar_2020}.
A market capitalization computed via the free float shares is called float adjusted. Stocks in the Russell US indexes are weighted by their float-adjusted market capitalization times the closing price of the corresponding pricing vehicle.

The \textit{rebalance day} is scheduled to be one month later than the ranking day, coinciding therefore with the end of June or beginning of July. On this day the new issues of the indexes officially replaces the previous ones in the stock market. 
Minor adjustments to the indexes  are made in the period between the ranking day and the rebalance day, for example, in the case of mergers and spin-off of companies.
Since the FTSE Russell acquisition of the Russell U.S. indexes, which took place in 2015, the exact reconstitution calendar have been published on the FTSE Russell webpage. In order to reliably retrieve the reconstitution calendar prior to  FTSE Russell acquisition one has to consider the research literature. Specifically, in Table \ref{table:comparison_rebalance} we gather the  rank days and rebalance days as described in Section  ``Index Construction and Sample Selection'' of \cite{Cai2008}, Section 2 of \cite{Chen2006} and Section 3.2 of \cite{Madhavan2001}. It is documented by the FTSE Russell webpage \cite{ftse_calendar_2017} that, starting from 2017, the Russell U.S. indexes rank day has seen a shift towards the first half of May, in agreement to what is also observed for the year 2020 in Table \ref{table:comparison_rebalance}.
The academic sources, that are dated before 2017, unanimously agree on the rank day coinciding with the 31$^{\text{st}}$ of March.
\begin{table*}[h!]
\centering
\begin{tabular}{ @{}p{5cm}p{4cm}p{4cm}p{1cm}@{} }
 \toprule
 \multicolumn{4}{c}{\textbf{Academic Sources and Annual Reconstitution Calendars }} \\
 \midrule 
\textbf{Source} & \textbf{Rank Day} & \textbf{Rebalance Day} & \textbf{Year} \\ \midrule
FTSE Russell \cite{ftse_calendar_2020}  & May 8 & June 26  & 2020 \\
\hline
Cai \& Houge \cite{Cai2008}  &  May 31   & June 30  &  2008  \\
 \hline
Madhavan \cite{Madhavan2001} &  May 31   &  July 1  & 2001  \\
 \hline
 Chen \cite{Chen2006} &  May 31   &  June 30   & 2006 \\
 \bottomrule
\end{tabular}
\vspace{\abovecaptionskip}
\caption{A comparison of the Russell indexes reconstitution calendar across different sources. }
\label{table:comparison_rebalance}
\end{table*}

In a similar fashion to the annual rebalance schedule, the ranking for inclusions of IPOs takes place at the end of Q3, Q4 and Q1. Approximately one month after each quarterly ranking date the index gets extended with the new eligible IPOs, as it can be seen from the 2019 quarterly rebalance calendar in Table \ref{tab-ipo_table}. As discussed in Section ``Defining Membership by size'' \cite{ftse_russell_2020}, the quarterly rebalance days are taken to be the third Fridays of September, December and March and the corresponding rank days are set to be 5 weeks before each quarterly rebalance day.

\begin{table*}[h!]
\centering
\begin{small}
\begin{tabular}{@{}p{4cm}p{3.4cm}p{3.4cm}p{3.4cm}@{}  }
\toprule
 \multicolumn{4}{c}{ \textbf{Russell U.S. Quarterly Rebalance Calendar 2019} }\\
\midrule
Quarterly additions & 2019-Q3 Additions &  2019-Q4 Additions & 2020-Q1 Additions  \\ \midrule
Initial offering period  & IPOs which initially price/trade between May 13 and Aug 16. & IPOs which initially price/trade between Aug 17 and Nov 15.  & IPOs which initially price/trade between Nov 16 and Feb 14. \\\hline
Rank date &16 Aug 2019   &15 Nov 2019 & 14 Feb  2020 \\
\hline
Rebalance date & 20 Sep 2019&  20 Dec 2019 &  20 Mar 2020 \\
\bottomrule
\end{tabular}
\end{small}
\vspace{\abovecaptionskip}
\caption{Quarterly IPO calendar for the 2019 Russell rebalance schedule. }
\label{tab-ipo_table}
\end{table*}

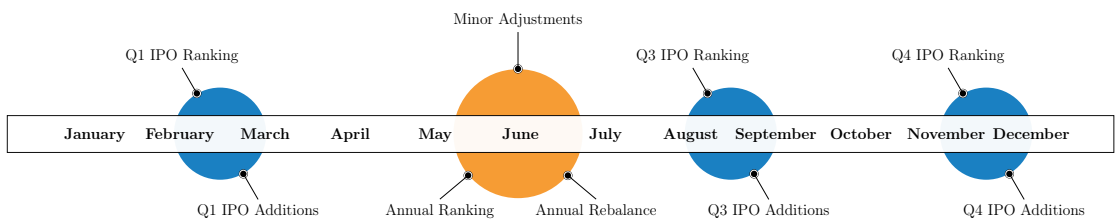
\begin{figure}[h!]
\centering
\resizebox{\textwidth}{!}{
\begin{tikzpicture}[timespan={}]
  \timeline[custom interval=true]{\textbf{January}, \textbf{February}, \textbf{March}, \textbf{April}, \textbf{May} , \textbf{June}, \textbf{July}, \textbf{August}, \textbf{September}, \textbf{October}, \textbf{November}, \textbf{December}}
  \begin{phases}
    \phase{between week=5 and 7 in 0.5,
      involvement degree=3.5cm,phase color=BurntOrange}
    \phase{between week=8 and 9 in 0.5,
      involvement degree=2.5cm,phase color=RoyalBlue}
    \phase{between week=11 and 12 in 0.5,
      involvement degree=2.5cm,phase color=RoyalBlue}
     \phase{between week=2 and 3 in 0.5,
      involvement degree=2.5cm,phase color=RoyalBlue}
  \end{phases}
  
  \addmilestone{at=phase-1.220,direction=220:1cm,
    text={Annual Ranking}, text options={below}}
  \addmilestone{at=phase-1.90,direction=90:1cm,
    text={Minor Adjustments}, text options={above}}
\addmilestone{at=phase-1.320,direction=320:1cm,
    text={Annual Rebalance}, text options={below}}
   \addmilestone{at=phase-4.120,direction=120:0.8cm,
     text={Q1 IPO Ranking}, text options={above}}
    \addmilestone{at=phase-4.300,direction=300:0.8cm,
     text={Q1 IPO Additions},text options={below}}
    \addmilestone{at=phase-2.120,direction=120:0.8cm,
     text={Q3 IPO Ranking}, text options={above}}
    \addmilestone{at=phase-2.300,direction=300:0.8cm,
     text={Q3 IPO Additions},text options={below}}
     \addmilestone{at=phase-3.120,direction=120:0.8cm,
     text={Q4 IPO Ranking}, text options={above}}
    \addmilestone{at=phase-3.300,direction=300:0.8cm,
     text={Q4 IPO Additions},text options={below}}
\end{tikzpicture}
}
\caption{Russell US indexes annual reconstitution timeline starting from June 2004. The timeline for the years 1989-2004 is identical apart from the quarterly IPOs additions, i.e. the blue circles.}
\label{fig:rebalance_schedule}
\end{figure}

The eligibility of the IPOs is established in two ways:
\\
\begin{enumerate}
\item If the new issue released in the IPO belongs to a company which is already an index constituent, the following criterion is considered. FTSE Russell determines the value associated to the IPO by multiplying the number of shares released in the IPO by the price of the pricing vehicle of the company releasing the issue. If the IPO's value is larger than the market capitalisation of the company sitting at the bottom of the Russell 3000E index, the security released in the IPO is added to the index. The market capitalisation of the company at the bottom of the Russell 3000E index, before being used in the comparison with the IPO's values, is suitably adjusted to take into account the price variations of the stocks which have taken place since the annual rebalance day. Note that the index membership assigned to the new issue will be the same of the pricing vehicle. However, the new issue is added to the index as a separate entity, therefore, it does not contribute to the total market capitalization of its company.
\item If the new issue belongs to a company which is \textit{not} in the index at the time of the IPO, then the market capitalisation of the IPO is established by multiplying the number of shares released, by their price on the quarter IPO ranking day. If such market capitalisation falls within any of the capitalisation breakpoints established at the annual ranking day\footnote{The capitalisation breakpoints are determined by the market capitalisation on the annual ranking day of the lowest ranking company in the Russell 1000 and 3000 index. For the Russell 2000 index they are similarly determined by the highest and lowest market capitalisation.} then the company is added to the one or more of the indexes accordingly.\\
\end{enumerate}
At every ranking day, either annual or quarterly, the index weights are recalculated based on the current capitalisation of the index constituents.
The current annual cycle of the indexes is  summarised by Fig. \ref{fig:rebalance_schedule}.
Note that once a company is delisted from the market, the corresponding securities are not traded anymore. This implies that if any such company is part of any Russell US index, then the number of actively traded securities in the index might reduce during the course of the year. Nonetheless, FTSE decides not to alter the current composition of the index and therefore any stock in the index which is delisted is not replaced.

Finally, as discussed in Section ``Long-Run Impact of Additions and Deletions'' of \cite{Cai2008}, the index returns can be found by a weighted average of the daily stock returns belonging to the index under the assumption of dividends reinvestment.
\section{The data} \label{sec-data}
\label{sec:crsp_data}
For the index reconstitution we will adopt the data available in the Wharton Research Data Services (WRDS) database, a research platform available to ``\textit{50,000 corporate, academic, government and nonprofit users at 400+ institutions in 30+ countries}'' provided by the the Wharton School of the University of Pennsylvania.
Specifically, we will limit ourselves to the financial data collected in the  Center for Research in Security Prices (CRSP) U.S. Stock database, which offers highly accurate information for the U.S. stock market. As further discussed in Section ``General Description: Coverage'' of \cite{crsp}, such database contains end-of-day and month-end prices for,
\vspace{+0.3cm}
\begin{itemize}
    \item \textit{NYSE}, starting from December 31, 1925, 
    \item  \textit{NYSE MKT}, starting from July 2, 1962, 
    \item \textit{NASDAQ}, starting from  December 14, 1972,
    \item \textit{Arca Exchanges}, starting from March 8, 2006.
\end{itemize}
\vspace{+0.3cm}
Moreover, the securities listed in this database are only equity securities for U.S. companies or international companies which are traded in any of the stock market aforementioned. The CRSP database contains the necessary information about the financial securities to be used in our analysis, such as prices, quote data, shares outstanding as well as the information about corporate actions, including IPOs. We refer the reader to Appendix \ref{app:crsp_details} for further technical details regarding CRSP databases and their content.
\section{Generating the Russell US indexes} \label{sec-res}
We turn to discuss the main features of our analysis to reconstruct Russell 1000, 2000 and 3000 using CRSP datasets to a very high degree of accuracy. Our analysis will not be free from approximations to the original Russell US index reconstitution, which currently counts more than 40 pages of methodology.
Due to the restricted breadth of data we consider, which as mentioned in Section \ref{sec-data} is confined to the CRSP U.S. stock financial data, our reconstitution methodology departs in multiple ways from the one of Section \ref{sec-meth}.  We will consider the time window starting from July 1989, the first year in which the annual rebalance schedule has been applied, and terminating in June 2019.

As already discussed in  Section \ref{sec-meth}, the exact annual reconstitution calendar of the Russell U.S. indexes is not available to the public for the entire time period we consider here. Therefore, we will take the annual rank day to take place on $\text{31}^{\text{st}}$ of May while the annual rebalance day to be the last Friday of June, as similarly supported by the academic sources cited in Table \ref{table:comparison_rebalance}. Following Table \ref{tab-ipo_table} and the methodology in \cite{ftse_russell_2020} we take the Q3, Q4 and Q1 rebalance days to fall on the third Friday of September, December and March respectively and the corresponding rank days to be 5 weeks theretofore.  If any rank day, be it annual or quarterly, falls on a U.S. non-trading day then we move it to the preceding trading day. Instead, for a rebalance day, be it annual or quarterly, which falls on a U.S. non-trading day we shift it to the following trading day. We do so in order to avoid any look-ahead bias. Such choice of schedule may deviate from the real one, but given that the rank and rebalance days often take place at the end of May and at the beginning of July respectively, we expect the deviation to be marginal and not to present any measurable effect on our final result.

Originally, as explained at length in Section \ref{sec-meth}, the pricing vehicle for each company is identified and then used to determine the market value of such company on the ranking day. Such procedure presents  extra work required for companies with more than one share class: we would need to identify the pricing vehicle using the two-year trading volume of each share class. Figure 2 of \cite{Cai2008} shows that it is possible to replicate, for the time period 1979-2004, to very high statistical accuracy the cumulative returns of the Russell 2000 index. This is done by computing the market capitalization of every company without determining its pricing vehicle, that is by multiplying the total shares outstanding of each security times the corresponding share price. This motivates us to deviate from the original methodology and to approximate the market capitalization of each company as in \cite{Cai2008}, for all the Russell U.S. indexes and for the entire time period 1989-2019. Moreover, we remark that such market capitalization would differ from the original one only for companies which are traded across two or more share classes and not for all the index constituents.

CRSP does not contain any information regarding cross-ownership or privately held shares. Such a piece of information is necessary in order to adjust for the free float, i.e. the fraction of shares which can be traded by the public. Hence, instead of computing the weights of the stock admitted to the index using the float-adjusted market capitalization, as in the original methodology, we use the the same market capitalization which was used to established the index ranking.

Starting from the reconstitution calendar of May 2004, we introduce quarterly ranking days and rebalance days in order to update our index with the IPOs taking place between rank days. Note that this is one of the main differences from previous index reconstruction papers such as  \cite{Cai2008}.  As already discussed in Section \ref{sec-meth}, the way  in which  the original methodology considers adding newly issued securities to the index is two-fold, depending whether they belong to a company listed in the index or not. Our methodology deviates from the original as follows. We will add only the newly issued securities belonging the companies not listed in the index. We require each security to satisfy the standard eligibility requirements for the admission to the index and whose IPO took place in the 3 months preceding the quarterly rank day. Once a new issues satisfies the eligibility requirements, then it can be added to one or more Russell U.S. indexes, only if its total market capitalization falls within the market capitalization breakpoints established during the most recent annual ranking day. 

Finally, in accordance with the original FTSE Russell methodology any company in the index which is deleted between rebalance days is never replaced.

Now we are ready to present our main results regarding indexes replication. We first concentrate on the results of indexes replication between 1989-2004, where Russell indexes were rebalanced annually, without any quarterly IPOs additions. Then we focus on more recent results of indexes replications between 2004-2019, where quarterly rebalancings including companies IPOs were introduced. 

\begin{figure}[h!]
\centering
\includegraphics[width=0.8\textwidth]{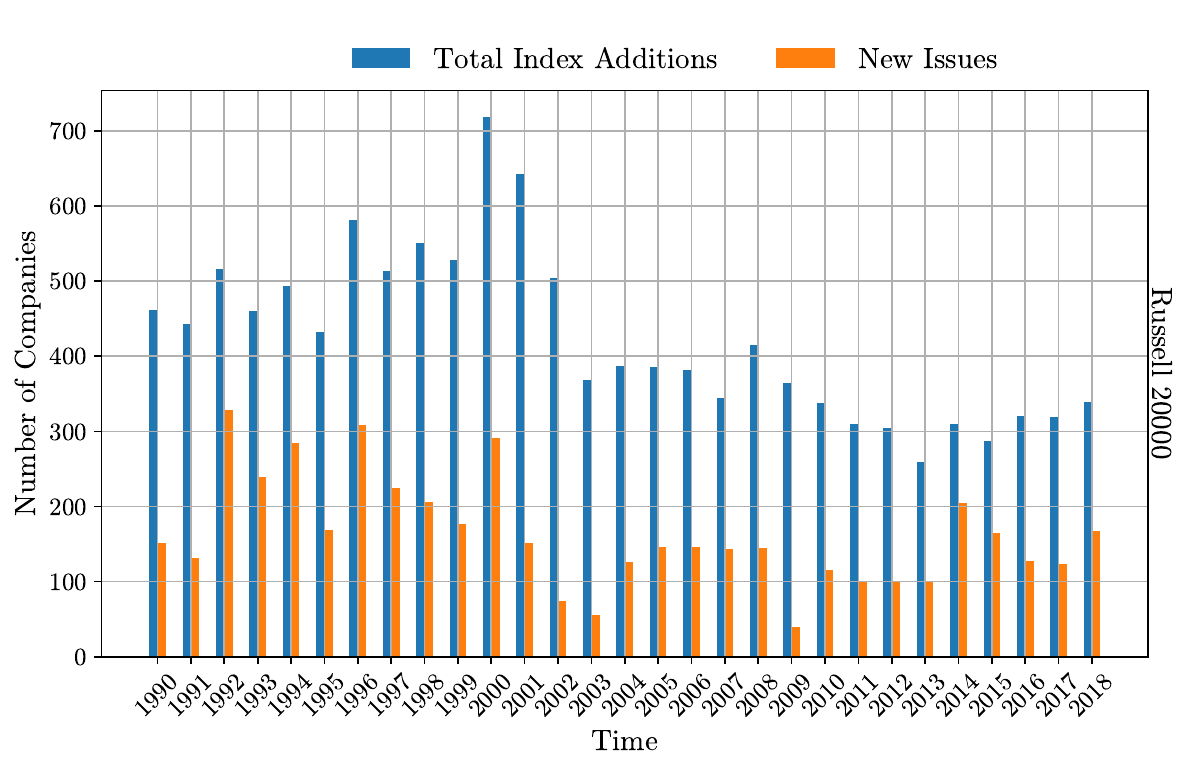}
\caption{The annual index changes between 1989-2019, as well the total number of new IPOs taking place in 12 months before year the rebalance of each year and satisfying the requirements for index additions. Russell US indexes started receiving index additions due to IPOs from September 2004.}
\label{fig:turnover_companies}
\end{figure}

Cai and Houge \cite{Cai2008} retrieved the roster of companies in the Russell 1000 and 2000 indexes for the time period 1979 to 2004 directly from Frank Russell Company. Figure 1 in \cite{Cai2008} displays the total number of Russell 2000 membership changes for each annual rank date alongside the number of new issues, i.e. IPOs and spin-offs, picked up by the index each year. In Fig. \ref{fig:turnover_companies} we also compute the annual number of constituent changes to Russell 2000 index for each annual reconstitution. Specifically, the \textit{``Total Index Additions''} bar at year $t$ counts the number of companies added to the Russell index during the year $t$ annual rebalance but which weren't in the index in the previous release of the  index. The \textit{``New Issues''} bar at year $t$ quantifies the companies added to the Russell 2000 index during the year $t$ annual rebalance whose IPO took place between May of year $t-1$ and May of year $t$. CRSP does not offer enough information regarding corporate actions in order to include spin-offs, as it was done in  Fig. 1 of \cite{Cai2008}. 
Over the years 1989 to 2004, where our methodology intersects with \cite{Cai2008}, we see that there is a very good agreement in terms of the number of new issues added to the index and the total number of index additions. Therefore, this guarantees that, for the time period 1989-2004, our methodology does not significantly differ from the original methodology which generated the rosters of companies studied by Cai and Houge and which were originally retrieved from Frank Russell Company. Figure \ref{fig:turnover_companies} extends the results of Cai and Houge to the time period in which IPOs were included in the original methodology, that is from the year 2004 until the most recent data.

We visually compare the index returns generated with our methodology versus the original index returns. We retrieve the original daily returns for the Russell 1000, 2000 and 3000  from the Bloomberg L.P. terminal. As already discussed, the index daily returns can be computed by a weighted average of the stock returns using the index weights. Therefore, a correct combination of the stocks selection and their corresponding index weights should be capable to reproduce the original daily index returns. We remark that it would be very hard, if not impossible, to back-engineer the constituents and the corresponding weights given the original daily index returns for the entire time period considered. The daily index returns are too noisy to be used for any meaningful visual comparison, therefore we plot the \textit{trailing three-months} (T3M) index gross returns. 
Let $0\leq t_{1}< t_{2}$ and let $r_{t}$  be the daily net return from day $t-1$ to time $t$. The compounded gross returns on $[t_{1},t_{2}]$ are given by,
\begin{equation*}
\label{eq:gross_returns}
 \prod_{t\in[t_{1},{t_{2}}]}(1+r_{t}).
\end{equation*}
Hence, the trailing three-months gross returns at time $t_{2}$ are computed by taking $[t_{1},t_{2}]$ to be a 3 months time window.

Figure \ref{fig:russell_returns} compares the T3M index gross returns based on the daily returns of our replicated indexes and the original Russell 1000, 2000 and 3000 indexes for the time window between July 1989 to June 2019. The Russell U.S. indexes generated following our methodology are consistently capable of mimicking the original Russell indexes for the entire duration of the time window considered.

The visual agreement of Fig. \ref{fig:russell_returns} is further assessed statistically via cross-correlations of  daily returns of our replicated indexes against the original Russell U.S. indexes. \newline
As discussed in Section ``What is the problem with cross-correlating simultaneous autocorrelated time series?'' of \cite{Dean2015}, the confidence bands of the cross-correlation between two time series has to be altered from the conventional cross-correlation limit, if the time series considered individually present significant autocorrelations. If such autocorrelations are not taken into account they may lead to the phenomenon of \textit{``spurious correlations''}, in which, as also shown in Fig. 1 of \cite{Dean2015}, where even two independent time series can present a significant correlation.

For the time window 1989-2004 and 2004-2019, we checked that none of the daily returns of the indexes, generated or original, present significant autocorrelation at any non-zero lag.  
The Pearson's correlations of generated and original indexes at zero lag are reported in Table \ref{table:correlations}, along with the 95\% confidence bands which are given by $\pm 1.96/\sqrt{n}$, where $n$ is the sample size. As a benchmark, in parentheses we report the cross-correlation at zero lag between a portfolio in which the components of the reconstructed Russell indexes are equally weighted and the original Russell indexes.
All the Pearson's cross-correlations are strongly significant. 

To further test the agreement among the series in terms of non linear measures of dependence, we also determine the Kendall's $\tau$ correlations between the reconstructed Russell indexes and the original ones. We report the values report in Table \ref{table:kendall-correlations} along with 95\% confidence bands which are given, for large samples, by $\pm 1.96 \sqrt{\frac{2(2n+5)}{9n(n-1)}}$ where $n$ is the sample size. Similarly to the case of Pearson's cross-correlations, we also report in parentheses the correlations between the equally weighted portfolio and the original Russell indexes. All the Kendall's $\tau$ correlations are strongly significant and outperform the benchmark of the equally weighted portfolio.  

\begin{table*}[h!]
\centering
\begin{small}
\begin{tabular}{ccccc}
\toprule
 \multicolumn{5}{c}{\textbf{Cross-Correlations (Pearson's)}} \\
\toprule
\textbf{Years} & \textbf{Russell 3000} & \textbf{Russell 2000} & \textbf{Russell 1000}  & \textbf{95\% Confidence Bands}\\ 
\midrule
1989-2004 & 0.98 (0.88) & 0.97 (0.96) &  0.98 (0.94) & $\pm$ 0.03 \\ 
\hline
2004-2019 & 0.99 (0.95) &  0.99 (0.99) & 0.99 (0.97) & $\pm$ 0.03 \\ 
\bottomrule
\end{tabular} 
\end{small}
\vspace{\abovecaptionskip}
\caption{Pearson's cross-correlation at lag 0 days between daily net returns for Russell 1000, 2000 and 3000 generated with our reconstitution procedure versus the original Russell indexes. The values in parenthesis are the cross-correlations between a portfolio in which the components of the reconstructed Russell indexes are equally weighted and the original Russell indexes.}
\label{table:correlations}
\end{table*}

\begin{table*}[h!]
\centering
\begin{small}
\begin{tabular}{ccccc}
\toprule
 \multicolumn{5}{c}{\textbf{Cross-Correlations (Kendall's $\tau$)}} \\
\toprule
\textbf{Years} & \textbf{Russell 3000} & \textbf{Russell 2000} & \textbf{Russell 1000}  & \textbf{95\% Confidence Bands}\\ 
\midrule
1989-2004 & 0.92 (0.67) & 0.88 (0.86) &  0.92 (0.77)  & $\pm$ 0.02 \\ 
\hline
2004-2019 & 0.96 (0.77) & 0.94 (0.89)  &  0.95 (0.83) & $\pm$ 0.02 \\ 
\bottomrule
\end{tabular} 
\end{small}
\vspace{\abovecaptionskip}
\caption{Kendall's $\tau$ cross-correlation at lag 0 days between daily net returns for Russell 1000, 2000 and 3000 generated with our reconstitution procedure versus the original Russell indexes. The values in parenthesis are the cross-correlations between a portfolio in which the components of the reconstructed Russell indexes are equally weighted and the original Russell indexes.}
\label{table:kendall-correlations}
\end{table*}

Similarly, as shown in Fig. \ref{fig:distribution_pre_2004}, the normalised distribution of the daily returns overlap to a very good degree between June 1989 to June 2004. The agreement is also confirmed by the corresponding Q-Q plot.

Next, we turn to the time window ranging from June 2004 to June 2019.
As already discussed at the beginning of this section, we consider adding to our index only securities issued by companies which are not listed in the index at the time of their IPO. This approximation allows us to exclude the extra work of considering different criteria for the securities belonging to companies already in the index. The full discussion of such criteria was given in Section \ref{sec-meth}. Figure \ref{fig:ranking_ipos} displays a justification for such approximation. We compare, for each year from 2004 to 2019, the number of new issues from companies which are not in the index, namely \textit{``New Issues not in Index''}, to the number of new issues from companies which belong to the Russell 1000 or 2000 indexes, that is the \textit{``New Issues from Russell 1000''} and \textit{``New Issues from Russell 2000''} bars.  Specifically, the \textit{``New Issues from Russell 1000''} bar at year $t$ quantifies the eligible securities issued by a company in the Russell 1000 index in the time window from year $t$ to year $t+1$. Similarly, mutatis mutandis, for the \textit{``New Issues from Russell 2000''} and the \textit{``New Issues not in Index''} bars. We remark that given the hierarchical structure  of the Russell 1000, 2000 and 3000 indexes the sum of the new issues from companies which belong to the Russell 1000 and 2000 indexes is simply the total number of new issues from companies in the Russell 3000 index. \newline
For the entire duration of our analysis the \textit{``New Issues not in Index''} IPOs are about two orders of magnitude larger than the \textit{``New Issues from Russell 1000 and Russell 2000''} IPOs combined. In many years there are no new issues belonging companies belonging to the indexes, for example as in 2007 or 2016. 

Similarly to the time period 1989-2004, we compare the T3M index gross returns of the reproduced Russell index against those of the original ones. Again, Figure  \ref{fig:russell_returns} compares the T3M cumulative returns arising from the Russell 1000, 2000 and 3000 index generated with our methodology versus the original indexes between June 2004 to June 2019. Similarly to the pre-2004 returns, our generated index can fully imitate the original Russell indexes returns.

Moreover, for the time window 2004-2019, the daily returns do not show any autocorrelation both for the original and generated time series. As contained in Table \ref{table:correlations}, the cross-correlation between the generated and original returns is extremely significant for both the Russell 1000, 2000 and 3000 indexes. 
\afterpage{
\begin{landscape}
 \begin{figure}
  \centering
  \includegraphics[width=1.4\textwidth]{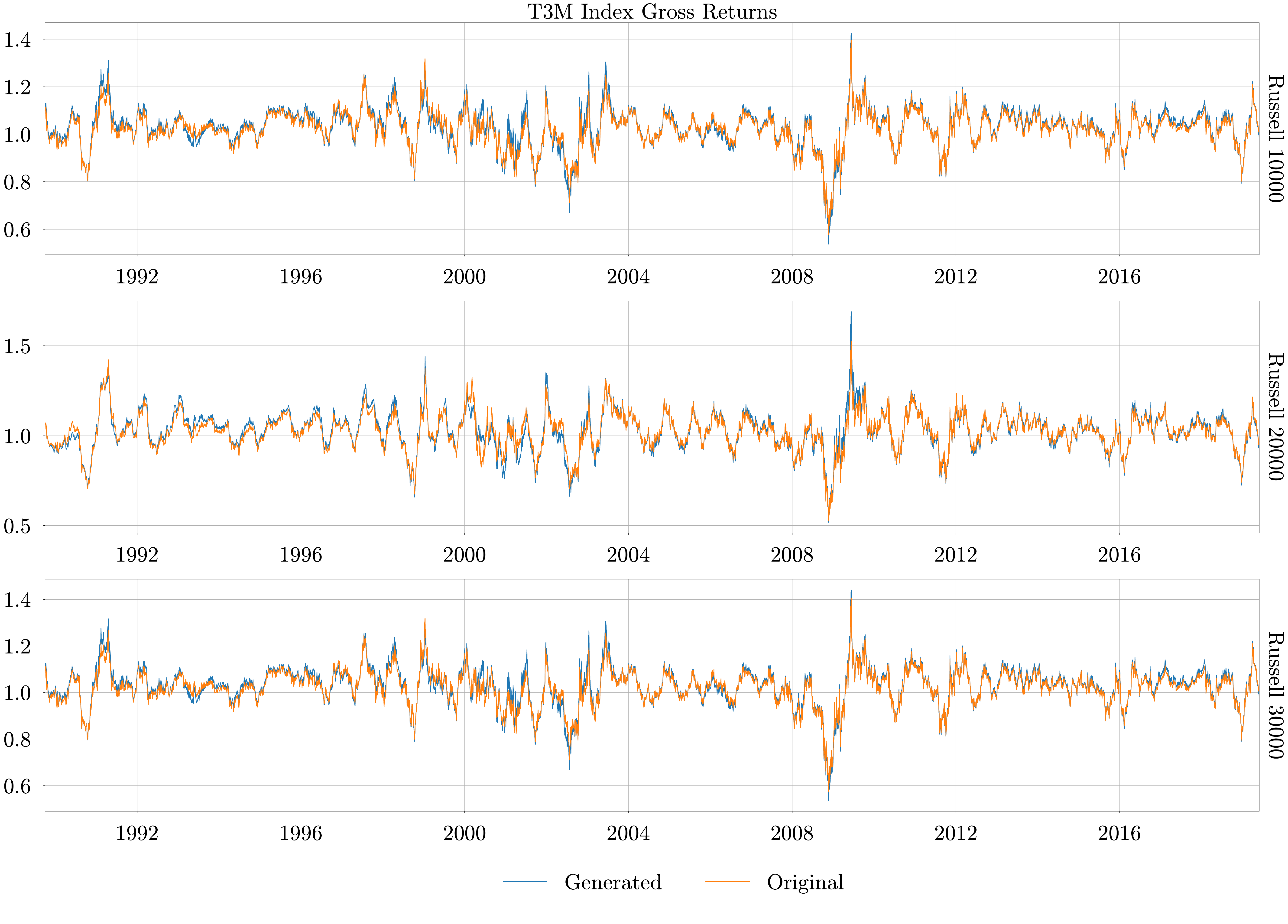}
  \caption{We compare the T3M gross returns for the time period June 1989 to June 2019 belonging Russell 1000, 2000 and 3000 generated with our reconstruction methodology (in blue) versus the original Russell US indexes (in orange). }
  \label{fig:russell_returns}
 \end{figure}
\end{landscape}
}

\afterpage{
\begin{figure}[h!]
\centering
\includegraphics[width=0.9\textwidth]{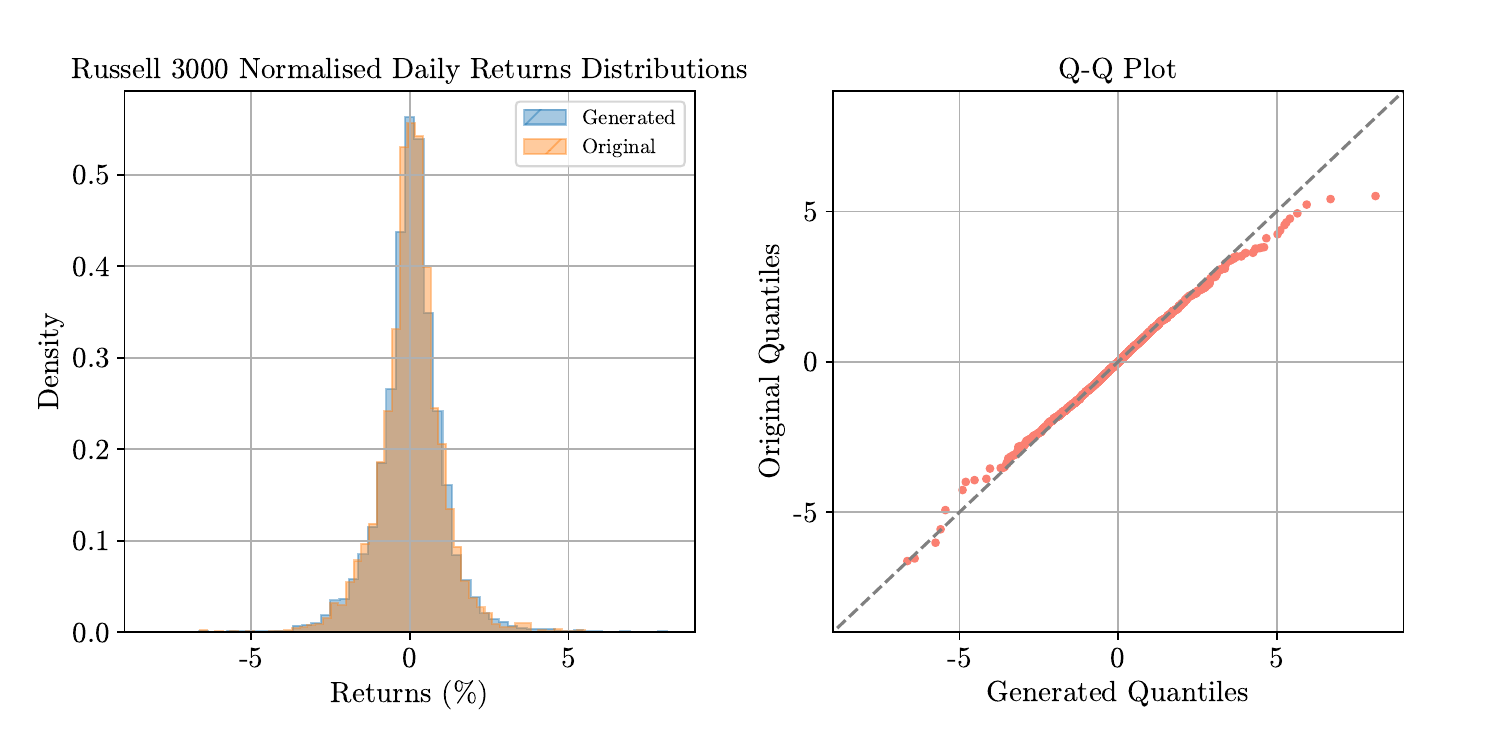}
\caption{On the left panel we compare between the normalised daily returns histogram of the Russell 3000 replicated index (orange area) and the original index (blue area), from June 1989 to June 2004. On the right panel the corresponding Q-Q plot between the two distributions is presented. }
\label{fig:distribution_pre_2004}
\end{figure}
}

\afterpage{
\begin{figure}[h!]
\centering
\includegraphics[width=0.9\textwidth]{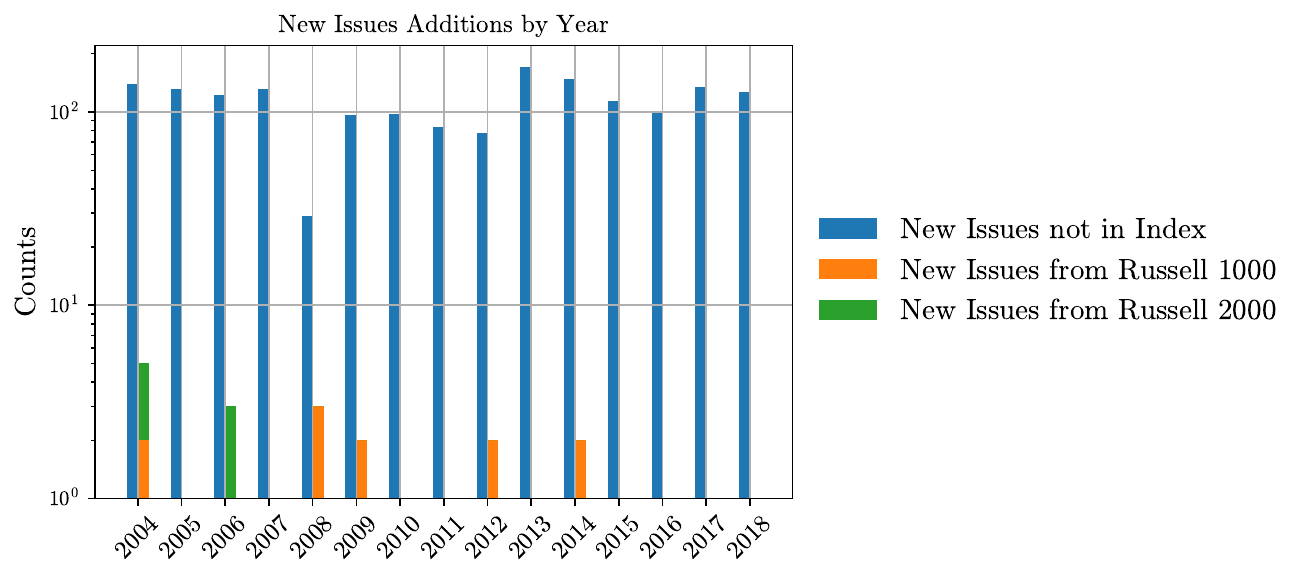}
\caption{Comparison of the number of new IPOs in each year in the following groups: \textit{``New Issues not in Index''} for securities which are not in the Russell 3000 index, \textit{``New Issues from Russell 1000''} and \textit{``New Issues from Russell 2000''} ,between 2004-2018.}
\label{fig:ranking_ipos}
\end{figure}
}

When comparing the normalised distributions of the daily returns as in Fig. \ref{fig:daily_returns_post_2004} we observe a very good agreement between our and the original indexes, which is supported by the respective the daily returns histogram (left panel) and a Q-Q plot (right panel) for the Russell 3000. 
\afterpage{
\begin{figure}[h!]
\centering
\includegraphics[width=0.9\textwidth]{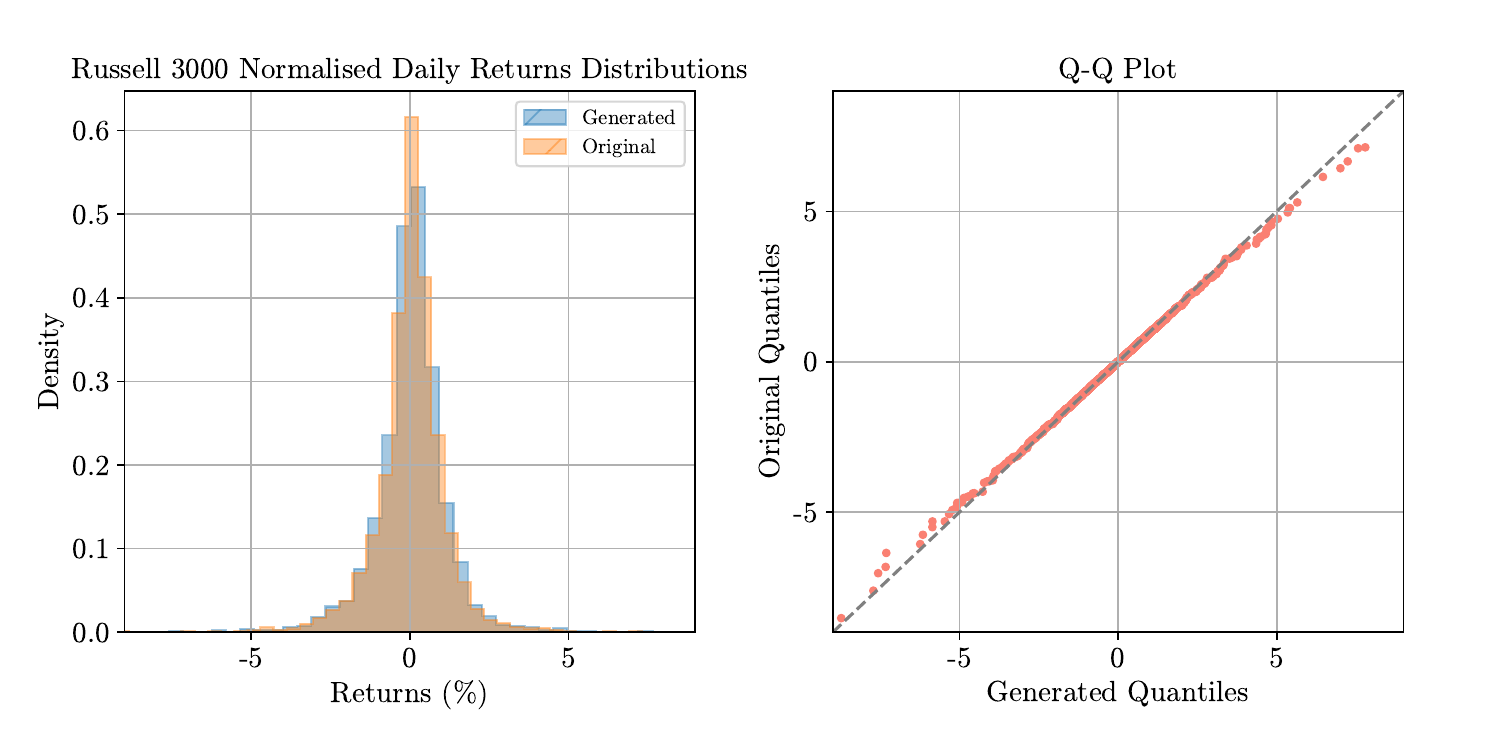}
\caption{On the left panel we compare between the normalised daily returns histogram of the Russell 3000 replicated index (orange area) and the original index (blue area), from June 2004 to June 2019. On the right panel the corresponding Q-Q plot between the two distributions is presented.}
\label{fig:daily_returns_post_2004}
\end{figure}
}

\section{Price Impact on Index Additions and Crowding}\label{sec-impact}
In this section we measure the temporary and permanent price impact for the annual additions and deletions in the Russell 3000 index. Moreover, we conduct a careful analysis on temporary and permanent price impact for new IPO's which are added to the the Russell 3000 index, estimating such quantities near the dates of quarterly and annual rebalancings. Studying the aforementioned price impact allows us to test 
whether the majority of market participants follow the index rebalance annually or quarterly. Specifically, for each year from 2004 to 2018 we test the following hypotheses:
\begin{itemize}
    \item  whether the most recent Q3, Q4 and Q1 quarterly additions remaining in the Russell 3000 index at annual rebalance, present a significantly different price impact compared to all the other additions in the index, near the date of the annual rebalance.
    \item whether near each of the Q3, Q4 and Q1 rebalancings, the quarterly additions have a price impact significantly different from other Russell 3000 index members, which have not changed their index membership in the most recent annual rebalance.
\end{itemize}

As a result we shed light on crowded and less crowded trades on new stock additions to the Russell 3000 index, in proximity of the annual and quarterly rebalance dates.

Madhavan, in his seminal work \cite{Madhavan2001},  measured the mean permanent and temporary price impacts generated by the annual addition and deletions of securities to the Russell U.S. indexes. Specifically, Madhavan focused on the 1996-2001 period, that is, before the index reconstitution methodology was updated to include the quarterly IPO additions as discussed in Section \ref{sec-meth}.  He computed the permanent and temporary price impact in terms of the log-returns produced by the securities within the following time intervals: for permanent impact, from the end of May until two months thereafter and for temporary impact from June 30 until one month thereafter. We recall that the end of May coincides with the annual rank day and June 30 can be considered the date of the reconstitution, as it can be inferred by Table \ref{table:comparison_rebalance}. It was found that for index additions over the period 1996-2001, the mean temporary impact and the mean permanent impact for the Russell 3000 index were 5.4\% and 3.3\%, respectively. For index deletions in the Russell 2000 index, the results were more modest with a mean temporary impact of 0.7\% and a mean permanent impact of –1.6\% (see Table 1 and 2 therein).

Quantifying temporary and permanent market impact is especially of interest to the market microstructure literature as well as to financial institutions since they are often found to be two of the main sources of transaction costs. 

Analogously to \cite{Madhavan2001}, we determine the permanent and temporary price impacts associated to the annual reconstitution of the index both for index additions and deletions.
We adopt the methodology from Section 5.2 of \cite{Madhavan2001} and quantify the temporary market impact as,
\begin{equation} \label{temp}
    R_{temp} =  \ln( p_{1} ) - \ln (p_{2})
\end{equation}
and the permanent market impact as,
\begin{equation} \label{per}
    R_{perm} = \ln (p_{2}) - \ln (p_{0}),
\end{equation}
where $p_{0}$, $p_{1}$ and $p_{2}$ are the stock prices at the annual rank day, one month thereafter and two months thereafter, respectively. Table \ref{table:impact_additions} reports the measurement for the permanent and temporary market impact for annual additions and deletions. The columns $\bar{R}_{temp}$ and $\bar{R}_{perm}$ contain the mean temporary and permanent market impact, respectively, expressed in terms of percentages with the corresponding standard errors in parenthesis. The column \textit{N\textsuperscript{\underline{o}}} contains the size of the sample considered. Note that we included in this analysis also the new IPOs which were added to the index on the annual rebalance, but not the ones that were added in the quarterly rebalancing of Q3, Q4 and Q1 of the same year.

We remark that after 2008,  the temporary market impact often presents a negative sign, presumably amenable to the 2010s bull market which signed a positive trend in the equity stock market, as also it has been documented by financial news e.g. \cite{wsj_2019,ft_2019,reuters_2019}. Nonetheless, in many years deletions still present a combination of positive temporary price impact and negative market impact regardless of the positive trend aforementioned.   Moreover, we also measure a more moderate price impact for deleted securities, analogously to what has been observed by \cite{Madhavan2001} for the time period 1996-2001.  A regression model can be applied in order to fully test the effect of market trends on $R_{perm}$ and $R_{temp}$. Since both the reference market index S\&P 500 and  individual stock monthly returns present significant autocorrelations across different lags \cite{JEGADEESH1990}, this type of analysis is quite involved and is left to a future work. 

\begin{table*}[h!]
\centering
\begin{tabular}{cccccccccccc}
\toprule
 & \multicolumn{11}{c}{\textbf{Price Impact in Russell 3000 Index}} \\
\midrule
 & \multicolumn{5}{c}{\textbf{Annual Additions}} & \multicolumn{5}{c}{\textbf{Annual Deletions}}\\
\midrule
Year & \multicolumn{2}{c}{$\bar{R}_{temp}$  }  & \multicolumn{2}{c}{$\bar{R}_{perm}$  }  & \textit{N\textsuperscript{\underline{o}}} & \multicolumn{2}{c}{$\bar{R}_{temp}$  }  & \multicolumn{2}{c}{$\bar{R}_{perm}$  }  & \textit{N\textsuperscript{\underline{o}}}  \\ 
\hline
2005 & -5.1 &(0.6)   & 9.7 &(1.1) & 344 &  -4.9 & (1.1) & 5.2& (1.8) & 242 \\ 
\hline
2006 & 4.6  &(0.7) & -10.7 &(1.0) & 348 & 2.0& (0.8)  & -7.1 &(1.1)& 232\\ 
\hline
2007 & 6.4 & (0.8) & -8.6 &(1.0) &  328 & 3.3& (0.7) & -3.2& (1.1) &  198 \\ 
\hline
2008 & 2.1 &(0.9) & -9.0 &(1.0) & 373  & 5.9 &(2.1) & -18.6 &(2.7) & 200\\ 
\hline
2009 & -1.6 &(0.9) & 8.8 &(1.3) &  302 &-5.2 &(1.6) & -0.6& (2.6) &   190\\ 
\hline
2010 & 5.6 &(0.8) & -9.3 &(1.3) &  296 & 2.0 &(1.3) & -10.8 &(1.9) & 200 \\ 
\hline
2011 & -0.7 &(0.7) & -5.1& (1.0) &  273 &  -1.0& (1.3) & -5.1 &(1.4) & 161\\ 
\hline
2012 & 1.5 &(0.8) & -1.5 &(1.1) &  265 & -1.1 &(1.9) & -6.6 & (2.4) & 148 \\ 
\hline
2013 & -4.7 &(0.8) & 6.1 &(1.1) & 236 & -3.2 &(1.6) & -4.5 & (2.9) & 121\\ 
\hline
2014 & 7.2 &(0.7) & -0.3 &(0.9) &   289 & 3.0 &(1.2)& 0.2 & (1.4) & 178\\ 
\hline
2015 & 4.5 &(1.2) & -4.2 &(1.6) &  254 & 12.8 &(1.9) & -16.5 & (2.4) & 162\\ 
\hline
2016 &  -5.4 &(0.6) & 1.0 &(1.2) &  283 & 0.2& (1.6) & -0.9 & (2.2) &  163\\ 
\hline
2017 & -0.8 &(0.8) &  1.8 &(1.2) &  285 &  1.8 &(1.1) & 1.0 & (1.8) &  156 \\ 
\hline 
2018 & -0.2 &(1.0) &  1.5 &(1.6)&  270 & 2.6 & (1.7) &  -5.8 & (2.2) & 155\\ 
\bottomrule
\end{tabular} 
\vspace{\abovecaptionskip}
\caption{A comparison of the mean permanent and temporary price impact for additions and deletions during the annual rebalance of the Russell 3000 index. The mean impacts are expressed in terms of percentages. In parenthesis the standard error of the mean.}
\label{table:impact_additions}
\end{table*}

As discussed in Section \ref{sec-meth}, from the 2004 annual reconstitution the Russell U.S. indexes has started receiving quarterly additions with newly issued securities in order to provide a version of the indexes which better resemble the equity market. We therefore investigate if such quarterly updates are really implemented by market participants via quarterly reconstitutions of the index portfolios. 

We recall that the price impact measured on the index additions arises from the transactions generated by traders portfolio rebalancings. In fact, close to the annual reconstitution period, market participants review their equity portfolios tracking the indexes: buy and sell orders are based on their beliefs on what constituents will be added and deleted from their current portfolio composition. It follows that the securities which are already present in the equity portfolio aforementioned at the time of the annual review  and are believed to remain in the new roster of securities, will not see an excess of transactions comparable to those of the new additions and deletions. Indeed, this is the reason why the index effect literature focuses exclusively on annual index additions and deletions. 

 In the hypothesis of a portfolio manager tracking the index at each quarter rebalance, at the time of the annual reconstitution she would mainly have to buy shares of the securities which she believes will be added to the index, and which were not added in any of the most recent Q3, Q4 and Q1 quarterly rebalancings. Contrarily, a portfolio manager tracking the index annually would need to buy the securities which she thinks are going to be added based on the previous annual index rebalance. Therefore, the latter portfolio manager may also buy securities that were added at the most recent Q3, Q4 and Q1 quarterly rebalancings, which are believed to remain in the index during the upcoming annual reconstitution.

If the majority of market participants were to update their index portfolios at each quarter, we would not expect to see significant market impact in those securities which were added at the most recent Q3, Q4 and Q1 rebalancings, and are believed to stay in the index in the upcoming annual reconstitution. On the other hand, one may consider the case where the majority of market participants update their index portfolios only at annual rebalancings. In this case we would expect that the securities which were added at the most recent Q3, Q4 and Q1 rebalancings to behave in a very similar fashion to any other security added to the index in the same year. In this section we are going to check  which of these cases applies in the market. 

Figure \ref{fig:violin_impact} compares the distribution of the permanent and temporary price impact for each annual reconstitution from 2005 to 2018, near the annual reconstitution dates. The top panel shows the distributions of the permanent price impact computed via \eqref{per}, while the bottom panel shows the distributions of the temporary market impact as defined in  \eqref{temp}.
For each year, the \textit{``Quarterly Additions''} group, which appears in orange, are the securities which were added at the most recent Q3, Q4 and Q1 rebalancings and which remained in the index in the upcoming annual reconstitution. The group \textit{``New Additions''}, in blue, represents any other security added to the index in the same year. We observe a very good agreement for the distributions at each year, supporting the hypothesis that securities in the \textit{``Quarterly Additions''} group and those in the  \textit{``New Additions''} group are traded by market participants in a very similar fashion within the time frame of up to two months after the annual reconstitution date.

We further investigate the observed similarity between the \textit{``Quarterly Additions''} group and the \textit{``New Additions''} group near the annual reconstitution date, under minimal assumptions. We conduct a two-sample \textit{t}-test assuming unequal variances and unequal sample sizes under the null hypothesis that the two groups are sampled from the same distribution. The $t$-statistic, which we denote by $t_{obs}$, is defined in \eqref{t-stat}. Here, $y$ refers to the log-returns of the  \textit{``New Additions''} group and $z$ stands for the returns of the \textit{``Quarterly Additions''} group according to \eqref{temp} and \eqref{per}.   We apply a bootstrap algorithm with 10,000 repetitions for each year. We refer to Algorithm \ref{alg-boot} in Appendix \ref{app:bootstrap} for the procedure used to calculate the $p$-values. In order to account for multiple hypothesis tests at each year from 2005 to 2018, we need to modify the $p$-values which are given by Algorithm \ref{alg-boot} by using the Benjamini-Hochberg procedure (see Section 3 of \cite{Benjamini1995}). In Appendix \ref{appen-multi} we describe the transformation that needs to be applied on the $p$-values (see equation (\ref{p-val-mod}) and Algorithm \ref{p-val-alg} therein). When adjusting for multiple testing, the \textit{p}-values for the hypothesis for the permanent price impact tests and those for the hypothesis temporary price impact tests are adjusted separately.

Table \ref{table:p-values} reports the two-tailed adjusted \textit{p}-values of our test statistic for each year from 2005 to 2018.  Only in one year out of fourteen, namely 2006, the mean permanent price impacts of the two groups were found to be significantly different at 0.05 significance level. As for the temporary market impact the two groups were found to be significantly different only on three years out of fourteen, namely 2006, 2011 and 2016. Nonetheless, such discrepancy could have already been deduced from Figure \ref{fig:violin_impact}, where the blue and orange temporary price impact distributions present visibly different features. 

The reconstitution methodology introduced in Section \ref{sec-res} allows us to keep track of the quarterly index additions in the Russell 1000, 2000 and 3000 indexes at each quarter. This allows us to test the complementary hypothesis of whether the new quarterly additions receive any abnormal price impact soon after the corresponding quarterly rank day. In fact, in the case in which most of market participants were to rebalance their index portfolio annually, the new quarterly additions would not see any significant excess of price impact compared, for example, to the securities which are already present in the index. 

As already discussed in this section, the new annual additions present an excess of price impact measurable up to the end of July of the corresponding year, i.e. two months after the annual rank day. Moreover, as shown in Table \ref{tab-ipo_table}, the Q3 rank day usually falls approximately in the middle of August. Hence, it might be the case that some new annual additions could continue to present a measurable excess of price impact in the proximity of Q3 rank day.
\afterpage{
\begin{landscape}
 \begin{figure}
  \centering
  \includegraphics[width=1.41\textwidth]{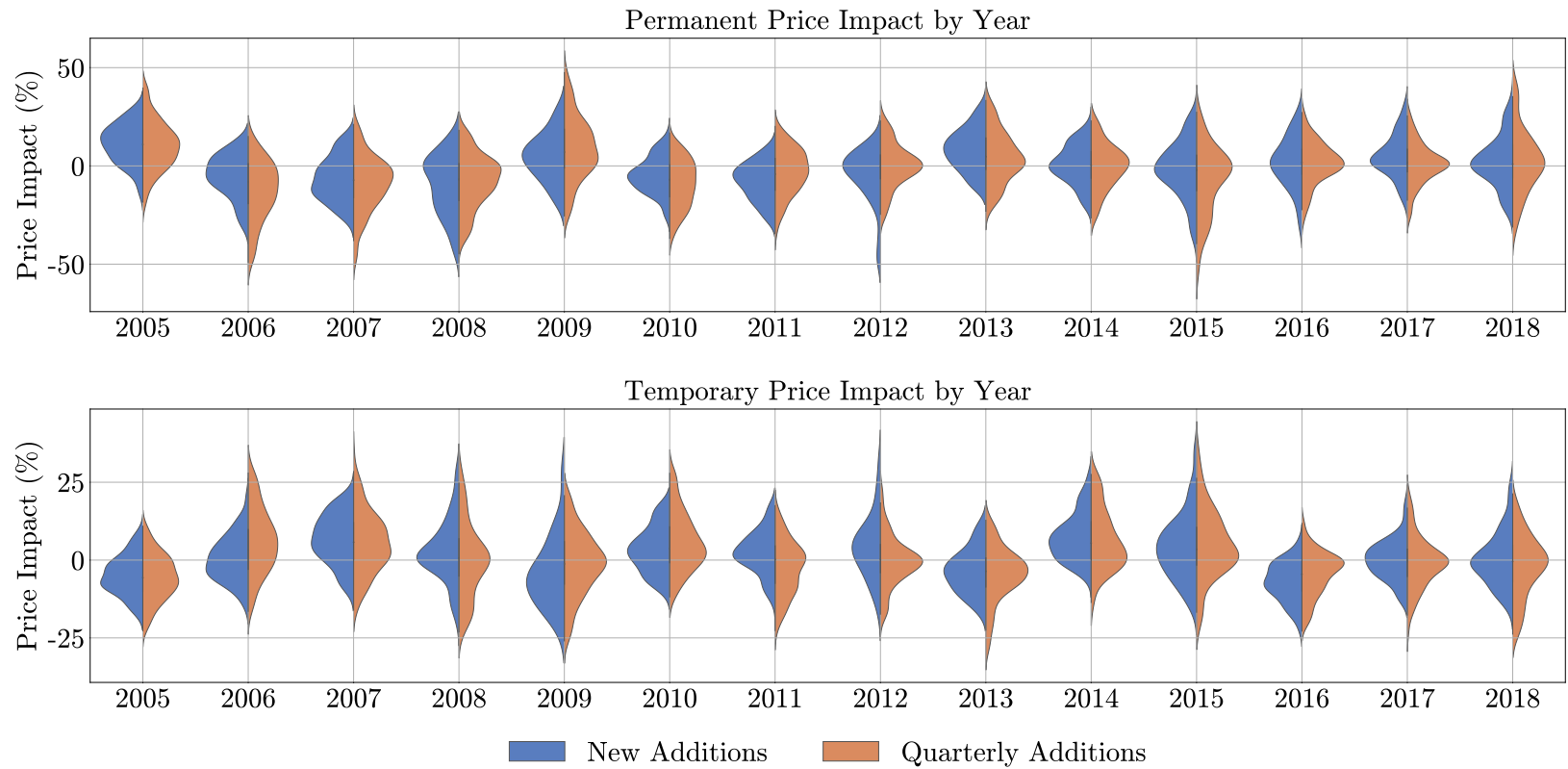}
  \caption{A comparison of the permanent and temporary impact distributions by year for the \textit{``Quarterly Additions''} group and the \textit{``New Additions''} group, for the Russell 3000 index. The \textit{``Quarterly Additions''} group are the securities which were added at the most recent Q3, Q4 and Q1 rebalancings and which remained in the index in the upcoming annual reconstitution. The \textit{``New Additions''} represent any other security added to the index in the same year.}
  \label{fig:violin_impact}
 \end{figure}
\end{landscape}}

The only securities in the index which can  be safely considered devoid of the aforementioned excess of price impact are those who have not changed their index membership in the most recent annual rebalance. In fact, even securities who remained in the Russell 3000 index during the most recent annual rebalance, but have moved from the Russell 2000 index to the Russell 1000 index, might still present an excess of price impact. This effect is generated by the buy orders of those market participants following the Russell 1000 index.
\afterpage{
\begin{table*}[h!]
\centering
\begin{small}
\begin{tabular}{ccccc}
\toprule
&\multicolumn{2}{c}{\textbf{Permanent Impact}} & \multicolumn{2}{c}{\textbf{Temporary Impact}} \\ 
\midrule
Years & $t_{obs}$ & $p$ &  $t_{obs}$ & $p$ \\ 
\midrule
2005 & -0.455 & 0.755 &  -1.261 &	0.322\\ 
\hline
2006 & -3.782 & 0.001 &  3.798 &	0.002\\ 
\hline
2007 & -1.246 & 0.340 & -0.326 &	0.870 \\ 
\hline
2008 & 1.186 & 0.340 &  -1.605 &	0.255 \\ 
\hline
2009 & 1.278 & 0.340 &  0.180 &	0.932 \\ 
\hline
2010 & -2.745 & 0.055 & 1.843	& 0.209 \\ 
\hline
2011 & 2.302 & 0.077 & -3.195 & 0.013 \\ 
\hline
2012 & 2.549 & 0.055 &  -2.431 & 0.055 \\ 
\hline
2013 & -1.490 & 0.327 &  -1.382 & 0.322 \\ 
\hline
2014 & -0.373 & 0.755 &  -1.273 & 0.322\\ 
\hline
2015 & -1.339 & 0.340 &  0.011 & 0.988 \\ 
\hline
2016 & 0.032 & 0.972 & 2.812 & 0.026 \\ 
\hline
2017 & -1.546 & 0.322 &  -0.778 & 0.553 \\ 
\hline
2018 & 0.644 & 0.653 &  -0.834 & 0.553 \\
\bottomrule
\end{tabular} 
\end{small}
\vspace{\abovecaptionskip}
\caption{Observed test statistic and the corresponding two-tailed \textit{p}-values at 0.05 level of significance, for each year from 2005 to 2018 under the null hypothesis that the \textit{``Quarterly Additions''} group and the \textit{``New Additions''} are sampled from the same distribution.}
\label{table:p-values}
\end{table*}
}
Therefore, we investigate if the price impact measured on the new quarterly additions and those securities that have not changed their index membership in the most recent annual rebalance present measurable differences. 

For each quarter we conduct, similarly to what we did for the annual rebalance, a two-samples \textit{t}-test assuming unequal variances and unequal sample sizes. Our \textit{t}-test  is done under the null hypothesis that the new quarterly additions and the securities that have not changed their index membership in the two most recent annual rebalancings  are sampled from the same distribution. Again, we take the 10,000 repetitions for the bootstrap resampling as in Algorithm \ref{alg-boot}.
\afterpage{
\begin{figure}[h!]
\centering
\includegraphics[width=\textwidth]{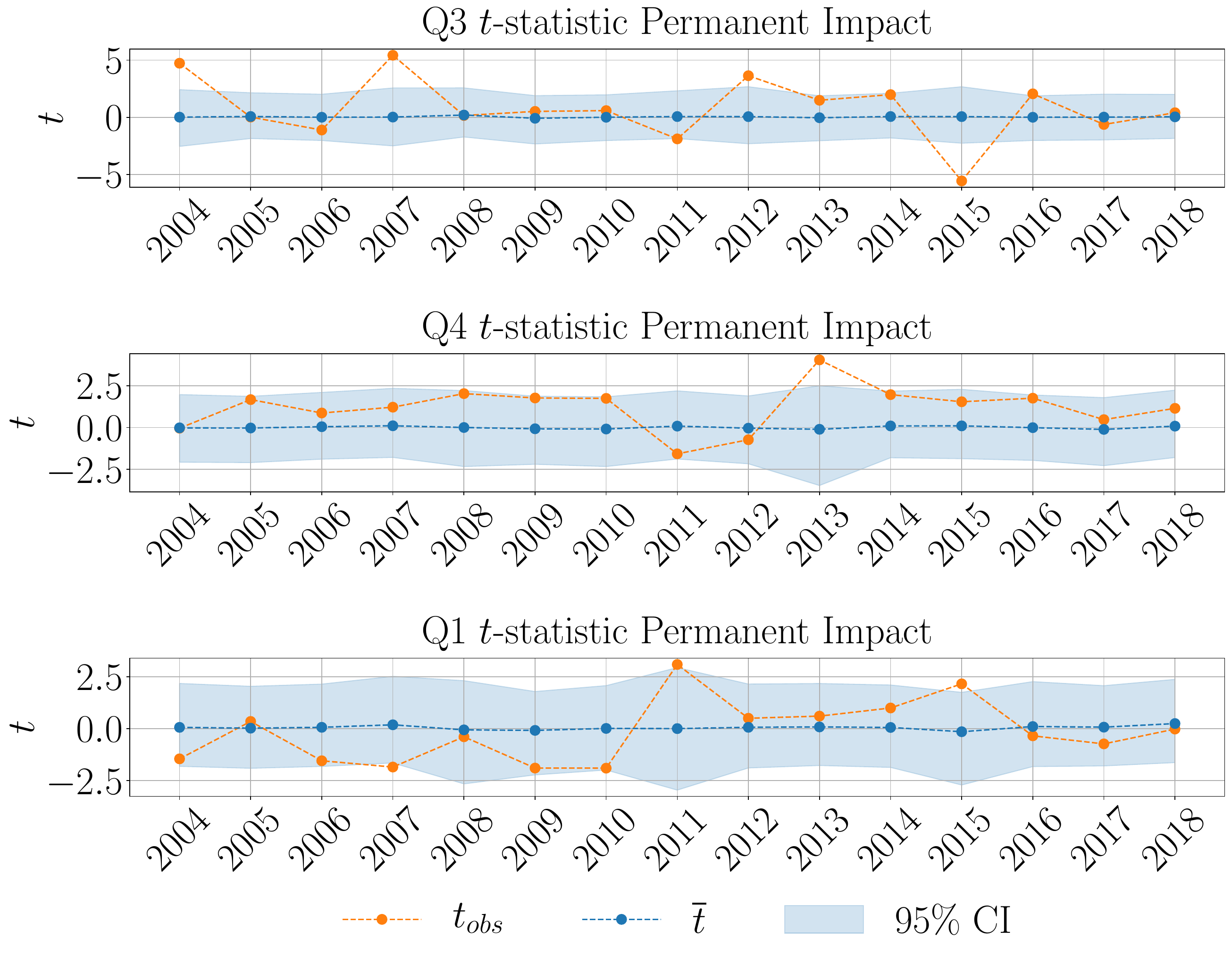}
\caption{ Permanent price impact case. The mean test statistic $\bar{t}$ is plotted in the blue line for the new quarterly additions versus those securities that have not changed their index membership in the most recent annual rebalance. The light blue bands are the 95\% confidence intervals for $t$-test statistic.  Observed test statistic $t_{obs}$ is presented in the orange line. }
\label{fig:test_quarterly_permanent}
\end{figure}
}

We define the mean test statistics $\bar{t}$ to be the mean of the bootstrap $t$-distribution created in Algorithm \ref{alg-boot},
for the two-samples \textit{t}-test, measuring permanent price impact. Here $y$ and $z$ refer to the two months log-returns starting on the quarter rank day, for stocks which are already in the index and for quarterly additions, respectively.

The 95\% confidence interval is also derived by using a bootstrap percentile method, as defined in \eqref{eq:percentile_boot}. 
The 95\% confidence interval needs to be further adjusted for multiple testing, as introduced by \cite{Benjamini2005} and further discussed in Section ``False Coverage Statement Rate-Adjusted CIs'' of \cite{Groppe2017}. This is done analogously to the $p$-value corrections of Table \ref{table:p-values}, see Algorithm \ref{alg-ci} in Appendix \ref{appen-multi} for the exact procedure. When adjusting for multiple testing, the 95\% confidence interval for the hypothesis for the Q3, Q4 and Q1 permanent price impact are adjusted separately.
Figure \ref{fig:test_quarterly_permanent} presents the mean test statistics $\bar{t}$ for the two-samples \textit{t}-test for permanent price impact (in the blue line), along with the 95\% confidence interval  (the light blue region).
In the orange line we show the observed test statistics $t_{obs}$ from \eqref{t-stat} 

Finding $t_{obs}$ outside the 95\% confidence interval would mean that we must reject the null hypothesis that the two samples come from the same distribution, and accept the alternative hypothesis that the distributions generating the two samples are different. We remark that only three years out of fifteen present two or more significant observed test statistics $t_{obs}$, namely 2007 and 2015. 

\begin{figure}[h!]
\centering
\includegraphics[width=\textwidth]{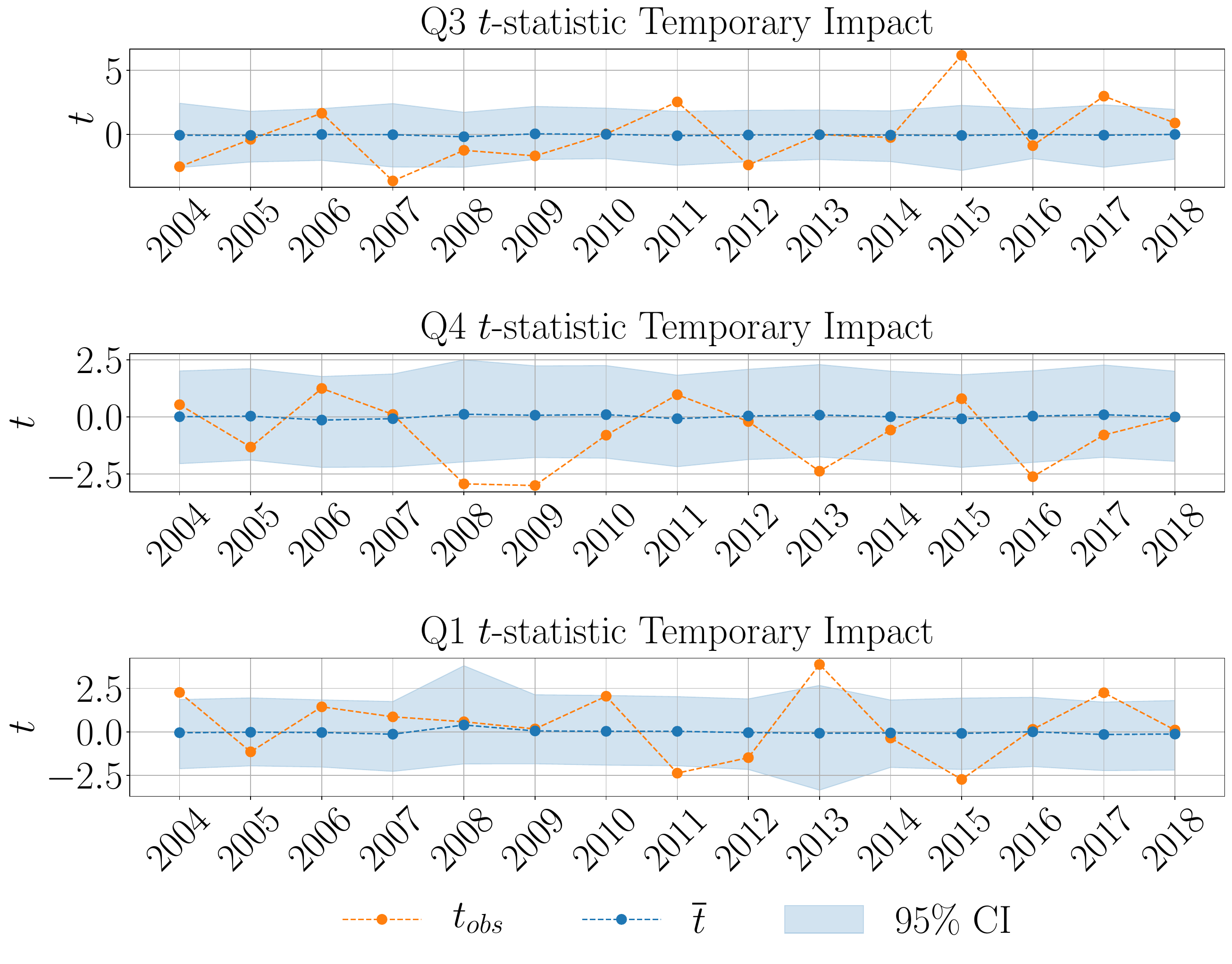}
\caption{ Temporary price impact case. The mean test statistic $\bar{t}$ is plotted in the blue line for the new quarterly additions versus those securities that have not changed their index membership in the most recent annual rebalance. The light blue bands are the 95\% confidence intervals for $t$-test statistic.  Observed test statistic $t_{obs}$ is presented in the orange line.}
\label{fig:test_quarterly_temporary}
\end{figure}

Similarly, in Figure \ref{fig:test_quarterly_temporary} we present in the blue line the mean test statistic $\bar{t}$ from Algorithm \ref{alg-boot}, for temporary price impact. Here $y$ and $z$ refer to the one months log-returns starting on the quarter rank day, of stocks which are already in the index and of quarterly additions, respectively. In the light-blue region we plot the $95\%$ confidence interval, which is derived along the same lines as in Figure  \ref{fig:test_quarterly_permanent}. In the orange line we show the observed test statistic $t_{obs}$ from \eqref{t-stat}, for  temporary price impact. We observe that only four years out of fifteen present two or more significant observed test statistics $t_{obs}$, namely 2011, 2013, 2015 and 2017. Nonetheless, the majority of the significant $t_{obs}$ are only marginally significant. It is reasonable to believe that the results in the years 2007 and 2009 might have been biased by the unfolding of the 2007-2008 financial crisis.

Ultimately, no compelling evidences were found to conclude that the majority of market participants follow the quarterly index rebalancings, as shown by Figures \ref{fig:test_quarterly_permanent} and \ref{fig:test_quarterly_temporary}. Moreover, the similarities between the price impact distributions observed in Figure \ref{fig:violin_impact} are in favour of the hypothesis that most market participants focus on the annual index rebalance, disregarding the quarterly index additions until the entire index portfolio has to be reviewed to take into account the changes brought by the annual index reconstitution.

The non-crowding phenomenon around quarterly rebalance dates, points out a possibility for profitable trading strategies on IPOs additions. A trader who wishes to track the new index additions could purchase new IPOs  additions around quarterly rebalance dates, with relatively low transaction costs. These IPOs could be sold later by the trader near the annual rebalance date, where the stock price will experience a significant increase due to price impact.

\section{Conclusion} \label{sec-con} 

This paper was built out of two parts.
On the first part we dealt with reconstruction of Russell US indexes.
We reviewed the main features of the Russell US index reconstitution methodology, starting from the index eligibility criteria to the quarterly IPOs addition procedure. Our analysis focused on the years 1989-2019 . We split our analysis into two time windows: the first is 1989-2004, and the second is 2004-2019, with 2004 being the year in which the quarterly IPOs addition were introduced. By a careful choice of approximations to the aforementioned methodology we reproduced the Russell 1000, 2000 and 3000 indexes to a very high degree of accuracy, using only CRSP US Stock database for our index reconstruction. We remark that the CRSP database, which is part of the Wharton Research Data Services (WRDS) database, is frequently used by researches in the field and is available in many academic institutions. 

The index constituents and their corresponding weights are released via a python package called \emph{pyndex} \cite{micheli_2020}, in the purpose to make this an accessible and standard platform for researches in the field, as the Russell indexes historical data is often unavailable for academic studies.   

In the second part of the paper, we studied crowding phenomenon on strategies that tracking the Russell 3000 index. 
We measured the temporary and permanent price impact for the annual index additions and deletions from 2005 to 2018.  We compared the permanent and temporary price impact affecting the securities added in the Q3, Q4 and Q1 quarterly rebalancings and remaining in the index versus the new index additions that didn't belong to the Russell 3000 index at any time in the previous rebalance year. Such measurements suggested a larger presence (or crowding) of trading strategies  that are tracking the index additions annually compared to those who rebalance quarterly. This phenomenon implies that indexing strategies can experience reduced transaction costs by buying new IPOs additions closely quarterly rebalance dates.

It was shown in \cite{volpati2020zooming} that common strategies, which are based only on momentum signals, are crowded and therefore would give a rather poor profitability.  Our finding add additional information on crowding phenomena, as we show that indexing strategies are indeed crowded on the one year scale but much less crowded on the $3$ months scale near quarterly rebalancings.

\section{Data availability statement}
Wharton Research Data Services (WRDS) was used in preparing this paper. This service and the data available thereon constitute valuable intellectual property and trade secrets of WRDS and/or its third-party suppliers.

\appendix
\section{CRSP US Financial Data}
\label{app:crsp_details}
The analysis in this paper is extensively based on the financial data available in the CRSP dataset. We summarise some of the main features of the databases considered. 

 The financial information regarding the securities used for this study was retrieved from the \texttt{CRSPQ:DSF} dataset, which consists of the quarterly updated CRSP daily stock data.
 The  \texttt{CRSPQ:DSF} dataset contains all the major daily financial indicators for the securities traded in the U.S. stock market, including stock closing prices, daily returns and outstanding shares. On the other hand, the \texttt{CRSPQ:DSFHDR} dataset contains the metadata related to each security in the \texttt{CRSPQ:DSF}. The information stored in the \texttt{CRSPQ:DSFHDR} file includes for example the initial day of trading for each security, the name of company to which the security belongs to as well as its Standard Industrial Classification.

Two labels are required in order to identify a company and its underlying securities. The identifier \texttt{permco} uniquely identifies a company in CRSP; it is neither reused once a company cease to exist, nor changed in case the company's name is subject to modification. Each company in CRSP may be traded on one or more securities therefore it is necessary to uniquely identify them in order, for example, to compute the company's market capitalisation. CRSP provides a unique five-digit permanent identifier for each security, under the name of \texttt{permno}, which neither changes during an issue's trading history, nor is reassigned after an issue ceases trading.
For each security, the daily returns, assuming dividend reinvestment, are given by the column \texttt{ret} in the \texttt{CRSPQ:DSF} file.

As discussed in Section ``Calculations'' of \cite{crsp}, the market capitalisation of a given company in CRSP can be found by computing,
\begin{equation*}
    \sum_{i}p^{i}_{t}\cdot v^{i}_{t},
\end{equation*}
where $p^{i}_{t}$ is the unadjusted close price of day $t$ of security $i$ and $v^{i}_{t}$ is the corresponding unadjusted number of its shares outstanding. The sum is taken over all the securities belonging to the given company. Here, $p^{i}_{t}$ and $v^{i}_{t}$ corresponds to the columns \texttt{prc} and \texttt{shrout} of the \texttt{CRSPQ:DSF} dataset, respectively.

The selection of the securities with prices greater than 1\$ discussed in Section \ref{sec-meth} is performed over the adjusted stock prices. Following Section ``Adjusting for Stock Splits and Other Corporate Actions'' of \cite{crsp} the adjusted stock price of securities $i$ is given by
$$ \frac{p^{i}_{t}}{\beta^{i}_{t}},$$
where $p^{i}_{t}$ is the unadjusted close price of security $i$ at day $t$ and $\beta^{i}_{t}$ is the corresponding price adjustment factor. Here $\beta^{i}_{t}$ is stored in column \texttt{cfacpr} of  \texttt{CRSPQ:DSF} dataset.
The share types in CRSP are identified through a two-digit code, named as \texttt{hshrcd}, describing the type of shares traded. The file \texttt{CRSPQ:DSFHDR} stores the \texttt{hshrcd} for all the financial securities belonging to the \texttt{CRSPQ:DSF} dataset. Following  Appendix A.2 in Chapter 3 of \cite{Fong2005}, common stocks are represented by a \texttt{hshrcd} equal to 10 or 11.

In order to establish which IPOs will be included in which quarterly addition, one has to consider the corresponding IPO date. In CRSP, the first day of trading corresponding to an IPO is stored in the \texttt{begdat} variable from \texttt{CRSPQ:DSFHDR} dataset. The most widely used database in IPO research is SDC Platinum from Thomson Financial, which is currently not available in WRDS or CRSP. As reported in \cite{crsp}, a comparison between SDC and \texttt{CRSPQ:DSFHDR} indicates that the first trading days agree in 76\% of cases. This confirms that the IPO dates in the \texttt{CRSPQ:DSFHDR} dataset can be reliably used for the index reconstruction.

\section{Bootstrap Two-Samples $t$-test}
\label{app:bootstrap}
In this section we summarise some useful results on bootstrap two samples $t$-test, which are taken from Chapter 16.2 of \cite {efron_bootstrap_book}. 

We consider two samples $\bm{z}$ and $\bm{y}$ of sizes $n$ and $m$, respectively, from possibly different probability distributions $F$ and $G$. We would like to test the null hypothesis $H_{0}: F=G$. Let  $\bm{x}$ be the collection of all the observations in $\bm{y}$ and $\bm{z}$. We test $H_{0}$ with the following two-samples unequal variance and unequal size statistic $t(\cdot)$,   
\begin{equation} \label{t-stat} 
 t_{obs}\equiv t(\bm{x})= \frac{\bar{z} - \bar{y}}{\sqrt{\bar{\sigma}_{1}^{2}/n +\bar{\sigma}_{2}^{2}/m }},
\end{equation}
with
\begin{equation*}
         \bar{\sigma}_{1}^{2} = \frac{1}{n-1} \sum_{i=1}^{n} (z_{i}-\bar{z})^{2}, \quad
         \bar{\sigma}_{2}^{2} = \frac{1}{m-1} \sum_{i=1}^{m} (y_{i}-\bar{y})^{2}, 
\end{equation*}
where $\bar{z}$ and $\bar y$ are the means of samples $z$ and $y$, respectively. Algorithm \ref{alg:bootstrap_alg} computes the bootstrap test statistic and the corresponding two-tailed $p$-values.
\begin{algorithm}[h!]
\caption{\textbf{Bootstrap test statistic for testing $F=G$}} \label{alg-boot} 
\begin{enumerate}
    \item Draw $N$ samples of size $n+m$ with replacement from $\bm{x}$. Call the first $n$ observations $\bm{z}^{*}$ and the remaining $m$ observations $\bm{y}^{*}$.
    \item Evaluate $t(\cdot)$ on each sample,
    \begin{equation*}
    \label{eq:test-stat-boot}
        t(\bm{x}^{*,k}) = \frac{\bar{z}^{*} - \bar{y}^{*}}{\sqrt{(\bar{\sigma}^{*}_{1})^{2}/n +(\bar{\sigma}^{*}_{2})^{2}/m }} , \quad k=1,2,\ldots N
    \end{equation*}
    where $\bar{\sigma}^{*}_{1}$ and $\bar{\sigma}^{*}_{2}$ are defined on $\bm{z}^{*}$ and $\bm{y}^{*}$ accordingly.
    \item Approximate two-tailed $p$-values by
    \begin{equation*}
    \label{eq:p-val-boot}
        \hat{p}_{boot} =  1 - \frac{\sum_{j=1}^{N} \bm{1}_{ \{-t_{obs} \leq t(\bm{x}^{*,j}) \leq t_{obs}\}}}{N}.
    \end{equation*}
\end{enumerate}
\label{alg:bootstrap_alg}
\end{algorithm}
In our analysis we take the number of bootstrap repetitions $N$ to be $10000$.

Moreover, as discussed in Chapter 13.3 of \cite{efron_bootstrap_book}, given a level of significance $\alpha$, the corresponding confidence interval for the bootstrapped distribution of the test statistic $t$ can be found using the \textit{bootstrap percentile method}. Let $\hat{\Phi}$ be the empirical cumulative distribution function of the bootstrap test statistic $t$. The $(1-\alpha)$ confidence interval are given by,
\begin{equation}
\label{eq:percentile_boot}
(\hat{\Phi}^{-1}(\alpha/2), \hat{\Phi}^{-1}(1-\alpha/2)), 
\end{equation}
where $\hat{\Phi}^{-1}(\alpha/2)$ and  $\hat{\Phi}^{-1}(1-\alpha/2)$ by definition correspond to the $\alpha/2$ and $1-\alpha/2$ percentiles, respectively.
\section{Multiple Testing} \label{appen-multi}
In this section we summarise some of the results regarding the Benjamini-Hochberg (BH) correction for independent multiple testing. 

As discussed in Section 2.b of \cite{rs_benjamini_yekutieli}, the \textit{p}-values can be adjusted for multiple testing according to the BH procedure via Algorithm \ref{p-val-alg}.
Section 3 of \cite{Benjamini1995} clarifies that in the BH procedure the test statistics are assumed to be independent.
\begin{algorithm}[h!]
\caption{\textbf{Multiple testing at significance level $\alpha$}} \label{p-val-alg}
Let $H_{0,i}$ with $i=1,\ldots,m$ be the null hypotheses, and $p_{i}$ be the corresponding $p$-values.
\begin{enumerate}
    \item Sort the $p$-values as $p_{(1)}\leq p_{(1)}\leq \ldots \leq p_{(m)}$ and let $p_{(k)}$ be the largest value such that 
    \begin{equation}\label{eq:p_values_bh}
    p_{(k)}\leq \frac{k\alpha}{m}
    \end{equation}
    \item If no such $k$ exists, select no discovery. Otherwise, reject the $k$ hypotheses corresponding to $p_{(1)}, \ldots, p_{(k)}$, declaring these findings to be discoveries.
\end{enumerate}
\label{alg:ci_alg}
\end{algorithm}
Let $H_{0,i}$ with $i=1,\ldots,m$, be the null hypotheses, and $p_{i}$ be the corresponding $p$-values. One can alternatively compute the BH-adjusted $p$-values as follows 
\begin{equation} \label{p-val-mod} 
   P_{(i)}^{BH} = \min \left(\left(\min_{j\geq i}mp_{j}/j\right),1\right).
\end{equation}
Then, $P_{(i)}^{BH}\leq \alpha$ if and only if $H_{(i)}$ is among the discoveries when using the BH procedure at significance level $\alpha$. 

As further discussed in Section ``False Coverage Statement Rate-Adjusted CIs'' of  \cite{Groppe2017}, the Benjamini-Hochberg procedure can be applied to confidence intervals for multiple comparisons as shown in the following algorithm. 
\begin{algorithm}[h!]
\caption{\textbf{Adjusted Confidence Intervals for Multiple-Testing }} \label{alg-ci}
\begin{enumerate}
    \item Apply the BH procedure to the $p$ values from the family of $m$ tests, where $m$ is the total number of hypothesis tests.
    \item For any $p$ value that is significant after the BH  procedure, construct a confidence interval for the corresponding test with coverage $1-\alpha'$ , where $\alpha'$ is:
    \begin{equation*}
    \alpha' = \left(\frac{k}{m}\right)\alpha,
    \end{equation*}
 with $k$ is defined as in (\ref{eq:p_values_bh}).
\end{enumerate}
\label{alg:ci_alg}
\end{algorithm}

In this paper we take $\alpha$ to be 0.05. Therefore, in order to compute the bootstrap confidence interval, adjusted to the Benjamini-Hochberg framework, we use $\alpha'$ in place of $\alpha$  in Eq. \eqref{eq:percentile_boot}.

\medskip 
 
\begin{thebibliography}{10}

\bibitem{Barroso2017}
P.~Barroso, R.~M. Edelen, and P.~Karehnke.
\newblock Institutional crowding and the moments of momentum.
\newblock SSRN.3045019, 2017.

\bibitem{Benjamini1995}
Y.~Benjamini and Y.~Hochberg.
\newblock Controlling the false discovery rate: A practical and powerful
  approach to multiple testing.
\newblock {\em Journal of the Royal Statistical Society: Series B
  (Methodological)}, 57(1):289--300, 1995.

\bibitem{Benjamini2005}
Y.~Benjamini and D.~Yekutieli.
\newblock False discovery rate--adjusted multiple confidence intervals for
  selected parameters.
\newblock {\em Journal of the American Statistical Association},
  100(469):71--81, 2005.

\bibitem{Biktimirov2004}
E.~N. Biktimirov, A.~R. Cowan, and B.~D. Jordan.
\newblock Do demand curves for small stocks slope down?
\newblock {\em Journal of Financial Research}, 27(2):161--178, 2004.

\bibitem{Boone2015}
A.~L. Boone and J.~T. White.
\newblock The effect of institutional ownership on firm transparency and
  information production.
\newblock {\em Journal of Financial Economics}, 117(3):508--533, 2015.

\bibitem{Bormetti2015}
G.~Bormetti, L.~M. Calcagnile, M.~Treccani, F.~Corsi, S.~Marmi, and F.~Lillo.
\newblock Modelling systemic price cojumps with {H}awkes factor models.
\newblock {\em Quantitative Finance}, 15(7):1137--1156, 2015.

\bibitem{bucci2019slow}
F.~Bucci, M.~Benzaquen, F.~Lillo, and J.~P. Bouchaud.
\newblock Slow decay of impact in equity markets: insights from the {AN}cerno
  database.
\newblock arXiv:1901.05332, 2019.

\bibitem{bucci2019trading}
F.~Bucci, F.~Lillo, J.~P. Bouchaud, and M.~Benzaquen.
\newblock Are trading invariants really invariant? {T}rading costs matter.
\newblock {\em Quantitative Finance}, pages 1--10, 2020.

\bibitem{Bucci2020}
F.~Bucci, I.~Mastromatteo, Z.~Eisler, F.~Lillo, J.~P. Bouchaud, and C.~A.
  Lehalle.
\newblock Co-impact: crowding effects in institutional trading activity.
\newblock {\em Quantitative Finance}, 20(2):193--205, 2020.

\bibitem{business_wire_2004}
BusinessWire.
\newblock Russell indexes to add ipos on a quarterly basis change in
  methodology enhances market representation of index.
\newblock Available at
  \url{https://www.businesswire.com/news/home/20040831005699/en/Russell-Indexes-Add-IPOs-Quarterly-Basis-Change},
  2004.

\bibitem{Caccioli2015}
F.~Caccioli, J.~D. Farmer, N.~Foti, and D.~Rockmore.
\newblock Overlapping portfolios, contagion, and financial stability.
\newblock {\em Journal of Economic Dynamics and Control}, 51:50--63, 2015.

\bibitem{Caccioli2014}
F.~Caccioli, M.~Shrestha, C.~Moore, and J.~D. Farmer.
\newblock Stability analysis of financial contagion due to overlapping
  portfolios.
\newblock {\em Journal of Banking \& Finance}, 46:233--245, 2014.

\bibitem{Cai2008}
J.~Cai and T.~Houge.
\newblock Long-term impact of {R}ussell 2000 index rebalancing.
\newblock {\em Financial Analysts Journal}, 64(4):76--91, 2008.

\bibitem{Lillo-Calcagnile}
 L.~M. Calcagnile, G.~Bormetti, M.~Treccani, S.~Marmi, and F.~Lillo.
\newblock Collective synchronization and high frequency systemic instabilities in financial markets.
\newblock {\em Quantitative Finance}, 18(2):237--247, 2018.

\bibitem{Capponi2019}
F.~Capponi and R.~Cont.
\newblock Trade duration, volatility and market impact.
\newblock SSRN.3351736, 2019.

\bibitem{Chang2014}
Y.~C. Chang, H.~Hong, and I.~Liskovich.
\newblock Regression discontinuity and the price effects of stock market
  indexing.
\newblock {\em The Review of Financial Studies}, 28(1):212--246, 2014.

\bibitem{Chen2005}
H.~Chen, G.~Noronha, and V.~Singal.
\newblock Index changes and unexpected losses to investors in {S}{\&}{P} 500
  and {R}ussell 2000 index funds.
\newblock SSRN.651950, 2005.

\bibitem{Chen2006}
H.~L. Chen.
\newblock On {R}ussell index reconstitution.
\newblock {\em Review of Quantitative Finance and Accounting}, 26(4):409--430,
  2006.

\bibitem{Cont2000}
R.~Cont and J.~P. Bouchaud.
\newblock {H}erd behavior and aggregate fluctuations in financial markets.
\newblock {\em Macroeconomic Dynamics}, 4(2):170--196, 2000.

\bibitem{Cremers2020}
M.~Cremers, A.~Pareek, and Z.~Sautner.
\newblock Short-{T}erm {I}nvestors, {L}ong-{T}erm {I}nvestments, and {F}irm
  {V}alue: {E}vidence from {R}ussell 2000 {I}ndex {I}nclusions.
\newblock {\em Management Science}, 2020.

\bibitem{ft_2019}
G.~Davies.
\newblock The great bull market reaches its 10th birthday.
\newblock Available at
  \url{https://www.ft.com/content/4f941406-5157-11e9-b401-8d9ef1626294}., 2019.

\bibitem{Dean2015}
R.~T. Dean and W.~T.~M. Dunsmuir.
\newblock Dangers and uses of cross-correlation in analyzing time series in
  perception, performance, movement, and neuroscience: The importance of
  constructing transfer function autoregressive models.
\newblock {\em Behavior Research Methods}, 48(2):783--802, 2016.

\bibitem{efron_bootstrap_book}
B.~Efron and R.J. Tibshirani.
\newblock {\em An Introduction to the Bootstrap}.
\newblock Chapman \& Hall/CRC Monographs on Statistics \& Applied Probability.
  Taylor \& Francis, 1994.

\bibitem{Fong2005}
H.~G. Fong.
\newblock {\em The World of Risk Management}.
\newblock World Scientific, 2005.

\bibitem{ftse_home}
FTSE-Russell.
\newblock Our story.
\newblock Available at \url{https://www.ftserussell.com/about-us/our-story}.

\bibitem{ftse_calendar_2017}
FTSE-Russell.
\newblock Russell {U.S.} indexes -- annual reconstitution.
\newblock Available at
  \url{https://www.ftserussell.com/press/ftse-russell-announces-schedule-annual-russell-us-index-reconstitution},
  2017.

\bibitem{ftse_russell_2020}
FTSE-Russell.
\newblock {R}ussell {U.S.} equity indexes v4.5 -- {C}onstruction and
  {M}ethodology.
\newblock Available at
  \url{https://research.ftserussell.com/products/downloads/Russell-US-indexes.pdf},
  2020.

\bibitem{ftse_calendar_2020}
FTSE-Russell.
\newblock Russell {U.S.} indexes review timetable - {M}arch 2020 and annual
  reconstitution timetable - {J}une 2020.
\newblock Available at
  \url{https://research.ftserussell.com/products/index-notices/home/getnotice/?id=2595207&_ga=2.179740323.629831544.1587596713-1989757251.1586952995},
  2020.

\bibitem{Groppe2017}
D.~M. Groppe.
\newblock Combating the scientific decline effect with confidence (intervals).
\newblock {\em Psychophysiology}, 54(1):139--145, 2017.

\bibitem{JEGADEESH1990}
N.~Jegadeesh.
\newblock Evidence of predictable behavior of security returns.
\newblock \emph{The Journal of Finance}, 45\penalty0 (3):\penalty0 881--898,
  1990.

\bibitem{Khandani2008}
A.~Khandani and A.~W. Lo.
\newblock What happened to the quants in august 2007?: Evidence from factors
  and transactions data.
\newblock SSRN.1288988, 2008.

\bibitem{Madhavan2001}
A.~Madhavan.
\newblock The {R}ussell reconstitution effect.
\newblock {\em Financial Analysts Journal}, 59(4):51--64, 2003.

\bibitem{micheli_2020}
A~Micheli.
\newblock pyndex - {R}ussell index reconstruction package.
\newblock Available at
  \url{https://github.com/alemicheli/pyndex}.


\bibitem{Onayev2008}
Z.~M. Onayev and V.~M. Zdorovtsov.
\newblock Predatory {T}rading {A}round {R}ussell {R}econstitution.
\newblock SSRN.1101341, 2008.

\bibitem{Petajisto2011}
A.~Petajisto.
\newblock The index premium and its hidden cost for index funds.
\newblock {\em Journal of Empirical Finance}, 18(2):271--288, 2011.

\bibitem{reuters_2019}
N.~Randewich.
\newblock Wall {S}treet's oldest-ever bull market turns 10 years old.
\newblock Available at
  \burl{https://uk.reuters.com/article/usa-stocks-bull/rpt-wall-streets-oldest-ever-bull-market-turns-10-years-old-idUKL1N20V1RJ},
  2019.

\bibitem{crsp}
Wharton Research~Data Services.
\newblock Overview of {CRSP} {U.S.} stock data.
\newblock Available at
  \url{https://wrds-www.wharton.upenn.edu/login/?next=/pages/support/data-overview/wrds-overview-crsp-us-stock-database/}.

\bibitem{wsj_2019}
Wall Street~Journal Staff.
\newblock Inside a {D}ecadelong {B}ull {R}un.
\newblock Available at
  \url{https://www.wsj.com/articles/inside-a-decade-long-bull-run-11552041001},
  2019.

\bibitem{volpati2020zooming}
V.~Volpati, M.~Benzaquen, Z.~Eisler, I.~Mastromatteo, B.~Toth, and J.~P.
  Bouchaud.
\newblock Zooming in on equity factor crowding.
\newblock arXiv:2001.04185, 2020.

\bibitem{rs_benjamini_yekutieli}
R.~Heller Y.~Benjamini and D.~Yekutieli.
\newblock Selective inference in complex research.
\newblock {\em Philosophical Transactions of the Royal Society A: Mathematical,
  Physical and Engineering Sciences}, 367(1906):4255--4271, 2009.

\bibitem{Zarinelli2015}
E.~Zarinelli, M.~Treccani, J.~D. Farmer, and F.~Lillo.
\newblock Beyond thesquare root: Evidence for logarithmic dependence of market
  impact on sizeand participation rate.
\newblock {\em Market Microstructure and Liquidity}, 01(02):1550004, 2015.

\end{thebibliography}
\end{document}